\documentclass[aps,reprint,superscriptaddress,amssymb,amsmath]{revtex4-2}
\usepackage{bm, siunitx, graphicx}
\usepackage[colorlinks=true,allcolors=blue]{hyperref}

\begin{document}

\title{Introducing fluctuation-driven order into density functional theory using the quantum order-by-disorder framework}
\author{Adam~H.~Walker}
\affiliation{London Centre for Nanotechnology and Department of Physics and Astronomy, University College London, London, WC1E 6BT, United Kingdom}
\author{Chris~J.~Pickard}
\affiliation{Department of Materials Science \& Metallurgy, University of Cambridge, 27 Charles Babbage Road, Cambridge CB3 0FS, United Kingdom}
\affiliation{Advanced Institute for Materials Research, Tohoku University, 2-1-1 Katahira, Aoba, Sendai 980-8577, Japan}
\author{Andrew~G.~Green}
\affiliation{London Centre for Nanotechnology and Department of Physics and Astronomy, University College London, London, WC1E 6BT, United Kingdom}

\begin{abstract}
Density functional theory in the local or semi-local density approximation is a powerful tool for materials simulation, yet it struggles in many cases to describe collective electronic order that is driven by electronic interactions.
In this work it is shown how arbitrary, fluctuation-driven electronic order may be introduced into density functional theory using the quantum order-by-disorder framework. This is a method of calculating the free energy correction due to collective spin and charge fluctuations about a state that hosts static order, in a self-consistent manner.
In practical terms, the quantum order-by-disorder method is applied to the Kohn-Sham auxiliary system of density functional theory to give an order-dependent correction to the exchange-correlation functional.
Calculation of fluctuation propagators within density functional theory renders the result fully first-principles.
Two types of order are considered as examples -- fluctuation-driven superconductivity and spin nematic order -- and implementation schemes are presented in each case.
\end{abstract}

\maketitle

\tableofcontents

\section{Introduction}

Density functional theory (DFT) \cite{Hohenberg1964, Kohn1965} is a powerful and efficient approach to modelling condensed matter systems and predicting new functional materials.
When studying correlated systems, one of the main challenges is to accurately describe collective electronic order.
Doing so successfully requires an ansatz that permits the order and an adequate description of the electronic interactions that drive it.
Spin-resolved DFT permits all charge and spin orders by describing a system using spin density matrices \cite{Barth1972}.
However, the most common approximations to exchange and correlation such as the local (spin) density approximation (LSDA) often fail to predict the correct ordered ground states of correlated materials.

An accurate description of correlation-driven order within DFT therefore requires improvements upon the local (or semi-local) density approximation to exchange and correlation.
For example, capturing non-collinear magnetic order such as spin density waves requires a more accurate description of the exchange energy \cite{Sharma2007,Kurth2009}, and electronically-driven nematic order and multipolar magnetic order are captured in LDA+U treatments  \cite{Long2020, Yamada2022, Mosca2022}.
In other cases, an improved description of electronic correlations is achieved by taking the Kohn-Sham eigenstates of DFT as a quasiparticle basis for many-body perturbative calculations.
This approach has been applied in $GW$ \cite{Aryasetiawan1992}, dynamical mean-field theory \cite{Anisimov1997}, 
spin-fluctuation \cite{Lischner2014,Lischner2015} and Eliashberg calculations \cite{Sanna2020}.
Superconductivity is precluded entirely by the spin density functional description, and instead requires the consideration of an additional Cooper pair density \cite{Lueders2005, Marques2005, Saunderson2020}.
			
In this work, a framework is presented in which arbitrary electronic order that is driven by collective charge and spin fluctuations is considered within density functional theory, in its local density approximation (LDA).
This is achieved using the quantum order-by-disorder (QOBD) approach to fluctuation-stabilised order \cite{Green2018, Conduit2009}.
In this method, collective order is introduced into an interacting system using a variational order parameter, and the effects of fluctuations in the ordered state are evaluated self-consistently within perturbation theory.
Here, the QOBD approach is applied to the Kohn-Sham auxiliary system of finite-temperature DFT. 
The result is an order-dependent correction to the exchange-correlation free energy functional.
By minimising the free energy of the ordered Kohn-Sham system, we find coupled equations for the order parameter and the Kohn-Sham eigenstates which must be solved in a self-consistent manner.

This framework allows the investigation of arbitrary order in correlated metals from first-principles. 
One of its key strengths is that multiple electronic orders may be considered simultaneously, within a single framework and on an equal footing. 
It is therefore suited to investigating the delicate competition and coexistence of electronic orders in correlated metals, such as the iron pnictides and chalcogenides \cite{Fernandes2022}, which can be achieved in a controlled manner. 
It also enjoys the advantages of an order parameter theory;
order can be considered in a particular channel \textit{via} the choice of ansatz, and the resulting optimisation problem is as simple as possible. 
In the case of spin order, for example, the optimisation of an order parameter is simpler than the optimisation over the full space of spin density matrices in spin-DFT.

The discussion is structured as follows.
Section \ref{section:QOBD} reviews the quantum order-by-disorder framework.
This is implemented in its field theoretical form using variational perturbation theory, which is a method of introducing collective order into an interacting system using a variational order parameter.
In order to put it on firm theoretical ground, a brief comparison is made to two equivalent ways to perform the QOBD calculation that use source fields and a Hubbard-Stratonovich decoupling.
In Section \ref{section:superconductivity}, the QOBD approach is illustrated using two examples: fluctuation-driven superconductivity and spin nematic order in the Hubbard model.
An analysis of the superconducting result shows that it can be viewed as a leading order, linearised limit of Eliashberg theory.
Section \ref{section:QOBD-DFT} then describes how fluctuation-driven order is introduced into spin-degenerate DFT by applying the QOBD method to the Kohn-Sham auxiliary system.
This is demonstrated separately for superconducting and spin nematic order, which are representative of orders in the particle-particle and particle-hole channels, and implementation schemes are presented in each case.
These results are rendered fully first-principles by calculating the charge and spin fluctuation propagators within DFT.
In the superconducting case, the QOBD-DFT  result is equivalent to the superconducting DFT that is formally derived \textit{via} a Cooper pair density \cite{Lueders2005, Marques2005}.

\section{\label{section:QOBD} Quantum order-by-disorder}

Quantum order-by-disorder \cite{Green2018, Conduit2009} is a framework in which to treat describe arbitrary collective electronic order that is stabilised by collective charge and spin fluctuations in correlated metals.
It is founded in the notion that the statics and dynamics of interacting quantum systems are inextricably linked.
In metallic systems, collective electronic order modifies the spectrum of spin and charge fluctuations by deforming the Fermi surface, which alters the phase space available for low energy particle-hole excitations. 
In turn, the correction to the free energy due to fluctuations may be sufficiently large so as to dictate the static order.
The collective order and the fluctuations it supports must therefore be treated self-consistently.
This is achieved  in QOBD by self-consistently evaluating the free energy of fluctuations about an ordered state.
Since an arbitrary number of electronic orders maybe considered simultaneously and on equal footing, this method excels when studying the phase diagram reconstruction in the vicinity of quantum critical points due to critical fluctuations \cite{Conduit2009, Karahasanovic2012, Conduit2013, Pedder2013, Hannappel2016}.
It has had success in describing the helimagnetic phase of MnSi \cite{Kruger2012}, the inhomogeneous magnetic phases of Sr$_3$Ru$_2$O$_7$ under applied field \cite{Berridge2009} and modulated magnetic phase of PrPtAl both in the presence and absence of applied field \cite{Abdul-Jabbar2015, O'Neill2021}. 

\subsection{The QOBD framework using variational perturbation theory}

The quantum order-by-disorder calculation can implemented by a number of routes. 
In this work it will be performed within a field theoretical framework using variational perturbation theory.
In this approach, collective order is introduced into a system of interacting electrons using variational terms, which are added and subtracted to the action of the system. 
These terms describe the coupling of electrons to a classical (i.e. static) field which behaves as a variational order parameter.
One set of variational terms is absorbed into the non-interacting action, and this breaks a particular symmetry.
The other set is treated alongside the electron-electron interaction as a perturbation on the ordered system.
This is treated up to second order, and is evaluated with respect to the ordered non-interacting action.
By only keeping the effects of the variational terms at leading order in the perturbative series, the order parameter dependence of the theory is non-cancelling.
The result is a theory in which the effects of spin and charge fluctuations about a static, ordered state are self-consistently evaluated.
Although variational perturbation theory can be implemented on the level of the Hamiltonian \cite{Conduit2013}, it is more convenient to work with the action when considering order in the particle-particle channel (e.g. superconductivity).

We will illustrate this approach by considering an arbitrary order in the Hubbard model of interacting electrons, with a Hamiltonian,
\begin{equation} 
\hat{H} = \sum_{\bm k\sigma} \left(\epsilon_{\bm k} -\mu \right) \hat n_{\bm k \sigma}  +  U \sum_{\bm r} \hat{n}_\uparrow(\bm r) \hat{n}_\downarrow(\bm r),
\label{eq:Hubbard_Hamiltonian}
\end{equation}
at fixed chemical potential, $\mu$.
We can write down the corresponding electronic partition function, $Z = \int D[c^\dagger, c] \text{exp} (-S[c^\dagger, c] )$, as an integral over configurations of Grassmann fields, $c,c^\dagger$.
Expressed in a frequency and momentum basis, the action is  
\begin{equation} \begin{aligned}
S[c^\dagger,c] 
&= \sum_{k\sigma} c^\dagger_{k\sigma} (-i\omega_n + \epsilon_{\bm k} - \mu)  c_{k\sigma} \\ 
&+  U \sum_{kpq} c^\dagger_{k+q \uparrow} c^\dagger_{p-q \downarrow} c_{p\downarrow}c_{k\uparrow} \\ 
&= S_{0} \ + \ S _{int},
\label{eq:Hubbard_action}
\end{aligned} \end{equation}
where $\omega_n = (2n+1)\pi T$ are fermionic Matsubara frequencies and $T$ is the temperature.
For clarity, we employ a 4-momentum notation, $k = (i\omega_n, \bm k)$.
We have also identified the non-interacting, $S_0$, and interacting, $S _{int}$, contributions to the action.
		
A desired collective order is introduced into the system using variational terms in the action, $S_{var}$, 
which describe the coupling of the electronic fields to a static collective field, $\phi$.
Their form is tailored to achieve a particular decoupling of the electronic interaction, and thus a particular static order.
For example, uniform magnetic ordering along the $z$ direction (whose order parameter is the magnetisation, $\phi = M$) couples to the imbalance in electronic spin densities, and therefore its variational terms are $S_{var}(M)[c^\dagger,c]  = U M \sum_{k\sigma} \sigma c^\dagger_{k \sigma} c_{k \sigma} $.
Similarly, spin-singlet superconducting order can be introduced by considering variational terms,
$S_{var}(\Delta)[c^\dagger,c]  =  -\sum_k ( \Delta_{\bm k} c^\dagger_{k\uparrow} c^\dagger_{-k\downarrow} + \bar{\Delta}_{\bm k} c_{-k\downarrow}c_{k\uparrow} )$, in terms of the gap function, $\Delta_{\bm k}$.
These are the terms that would result from a mean-field decoupling of the interaction in the spin or Cooper channels respectively.

The variational terms are added to and subtracted from the action, and then absorbed into its non-interacting and interacting parts,
\begin{equation} \begin{aligned}
S &= \underbrace{\vphantom{ \frac{1}{2} } S_0 - S_{var}} 
+ \underbrace{\vphantom{ \frac{1}{2} }S_{var} + S_{int}} \\ 
&= \ \ \ \ \ S_0^\prime \ \ \ \ \  + \ \ \ \ \ S_{int}^\prime,
\label{eq:variational_action}
\end{aligned}\end{equation}
and the partition function is rewritten as
\begin{equation} \begin{aligned}
Z[\phi] = \int D[c^\dagger,c]  \ e^{-S_0^\prime} \ e^{-S_{int}^\prime}.
\end{aligned}\end{equation}
The redefined interacting action, $S_{int}^\prime$, describes both the electron-electron coupling and the coupling of the electrons to the collective field, and it is treated perturbatively to second order.
Expanding the partition function \textit{via} the cumulant expansion, we find 
\begin{equation} \begin{aligned}
Z[\phi] &= Z_0[\phi] \ \langle e^{-S_{int}^\prime} \rangle_0 \\ 
&= Z_0[\phi] \ e^{ -\langle S_{int}^\prime \rangle_{0,c} \ + \ \frac{1}{2} \langle S_{int}^{\prime \ 2} \rangle_{0,c} \ + \ ... } \\ 
&=  e^{-S_{\text{eff}}[\phi]},
\label{eq:expanded_part}
\end{aligned}\end{equation}
from which we identify the effective action, $S_{\text{eff}}[\phi]$, that describes the collective field, $\phi$. 
$\langle ... \rangle_{0,c}$ denotes the connected contributions to the thermal average with respect to the non-interacting system, with action $S_0^\prime$, which self-consistently accounts for the effects of the order.
Since the variational terms are quadratic in the electron fields, the partition function of the non-interacting system, $Z_0[\phi]$, can be evaluated as
\begin{equation} \begin{aligned}
Z_0 [\phi]= \int D[\bar{c}, c]  e^{-S_0^\prime} =  e^{\text{Tr} \ln G_0^{-1}[\phi]}.
\end{aligned}\end{equation}
The expansion in Eq.~\ref{eq:expanded_part} is then curtailed at second order, leaving
\begin{equation} \begin{aligned}
S_{\text{eff}}[\phi]
&= \underbrace{-\text{Tr} \ln G_0^{-1}[\phi] \ + \ \langle S_{int}^\prime \rangle_{0,c} \phantom{\frac{1}{2}}  } \
\underbrace{- \ \frac{1}{2} \langle S_{int}^{\prime \ 2} \rangle_{0,c} } \\ 
& = \quad\quad\quad\quad\quad \ \ S_{_{SO}}  \quad\quad\quad \ + \quad\quad\quad  S_{fl}.
\label{eq:action_expansion}
\end{aligned}\end{equation}
The terms up to $O(S^\prime_{int})$  have been identified as the action of the static ordered (SO) system, and the $O(S_{int}^{\prime 2})$ term is the correction due to fluctuations (fl).
When evaluating the fluctuation contribution, we choose to neglect the variational part of $S_{int}^{\prime}$, i.e.
\begin{equation} \begin{aligned}
S_{fl}[\phi] = -\frac{1}{2} \langle S_{int}^{2}[\phi] \rangle_{0,c}.
\end{aligned}\end{equation}
If the contributions from $S_{var}$ were included in this term, the extra terms would cancel important order parameter-dependent terms in $S_{_{SO}}$. 
This is made clear by considering the full expansion of the effective action in Eq.~\ref{eq:expanded_part}: if this series was summed up, we would find that all dependence of the effective action on $\phi$ would cancel to zero (as in the original theory, Eq. \ref{eq:variational_action}).
		 	
The free energy is evaluated as $F[\phi] = -T \ln Z[\phi]$, giving
\begin{equation}
F   = F_{_{SO}} +  F_{fl},
\label{eq:QOBD_free_energy}
\end{equation}
where 
\begin{equation} \begin{aligned}
F_{_{SO}} &=  - T \ \text{Tr} \ln G_0^{-1}[\phi]  +  T \langle S^\prime_{int} [\phi] \rangle_{0,c}, \\ 
F_{fl} &= - \frac{T}{2} \langle S_{int}^{2} [\phi] \rangle_{0,c}. 
\end{aligned} \end{equation}
The thermal expectations are evaluated \textit{via} Wick contractions of the electronic fields, which give products of propagators of the ordered system, $G_{0}[\phi]$.
In this way $\phi$ enters the free energy of fluctuations.
The equilibrium value of $\phi$ is found by minimising the free energy, Eq.~\ref{eq:QOBD_free_energy}.
This differs from a conventional saddle point calculation in which the extremum of $F_{_{SO}}$ is located, giving the mean-field solution, $\phi_{_{MF}}$, and $F_{fl}[\phi_{_{MF}}]$ is a fluctuation correction. Here, the static and fluctuation contributions to the free energy both dictate the equilibrium order parameter value.
		
When considering multiple orders, one need only introduce multiple sets of variational terms.
The calculation then proceeds as above, only the electronic correlator now depends on multiple order parameters.
For example, in order to consider uniform charge and ferromagnetic orders, we decouple in the spin and charge channels using the variational terms
\begin{equation} \begin{aligned}
S_{var}(N, M) 
= U M  \ \sum_{k\sigma} \sigma c^\dagger_{k\sigma} c_{k\sigma} - U N  \ \sum_{k\sigma} c^\dagger_{k\sigma} c_{k\sigma},
\end{aligned} \end{equation}
where $N$ and $M$ are spatially uniform, classical charge and spin fields respectively.
Following the process outlined above gives the free energy,
\begin{equation} \begin{aligned}
F(N, M)  
= &- T \  \sum_{\bm k \sigma} \ln \left( 1 + e^{-\xi_{\bm k \sigma}(N, M)  /T} \right) \\
&+ U \Big( M \sum_\sigma \sigma n_\sigma  - N \sum_\sigma n_\sigma +  n_\uparrow n_\downarrow \Big) \\
&-2 U^2 \sum_{\bm k\bm p\bm q} 
\frac{ n_{\bm k+\bm q \uparrow} n_{\bm p- \bm q \downarrow} (1- n_{\bm p \downarrow}) (1- n_{\bm k \uparrow})}
{\xi_{\bm k + \bm q \uparrow} + \xi_{\bm p - \bm q \downarrow} - \xi_{\bm p \downarrow}  - \xi_{\bm k\uparrow}},
\label{eq:QOBD_ferro}
\end{aligned} \end{equation}
where $\xi_{\bm k \sigma}(N,M) =\epsilon_{\bm k} - \mu + U (N - \sigma M)$ is the dispersion of ordered state, measured relative to the chemical potential.
We have employed the compact notation, 
$n_\sigma \equiv \sum_{ k} \langle c^\dagger _{k \sigma} c _{k \sigma} \rangle = \sum_{\bm k} n_{\bm k \sigma} $ 
and $n_{\bm k \sigma} \equiv n(\xi_{\bm k \sigma}(N,M))$, where $n(\epsilon) = 1/(e^{-\epsilon/T}+1)$ is the Fermi function.
The saddle point of the $O(U)$ free energy retrieves the Stoner mean-field result,
$M = \sum_{\sigma} \sigma  n_{\sigma}/2  $ and 
$N =  \sum_{\sigma}  n_{\sigma}/2$.
There are some comments to make about the ferromagnetic result, Eq.~\ref{eq:QOBD_ferro}.
Firstly, the integral of the free energy of fluctuations contains a divergence that is a result of the contact interaction, 				which can be eliminated by a one-loop renormalisation of the Hubbard interaction vertex  \cite{Abrikosov1975, Karahasanovic2012}.	Secondly, the result makes clear the self-consistency of statics and dynamics:
the free energy of fluctuations depends on the static order \textit{via} the electronic dispersion, $\xi_{\bm k \sigma}$, but also dictates the equilibrium order parameter value.

\subsection{Equivalent implementation schemes}

The QOBD approach can be implemented by two equivalent methods, which use source fields \cite{Fukuda1995} and a Hubbard-Stratonovich transformation \cite{Karahasanovic2012}.
Here, brief comparisons will be made between these methods in order to put the variational perturbation theory approach on firm theoretical footing.

In the method of source fields, collective order is introduced into a field theory using source terms in the action, which describe the coupling of electrons to a  source field -- a fictitious field that breaks a particular symmetry.
In the case of ferromagnetic order along the $z$ direction, for example, this is a uniform magnetic field, $h$, which is conjugate to the magnetisation field, $M=\langle S_z \rangle$, where $S_z = \sum_{k\sigma} \sigma c^\dagger_{k \sigma} c_{k \sigma}$.
The source term is then, $S_{\text{source}} = h S_z$.
The free energy, expressed as a function of the order parameter, $F(M)$, may be determined by first finding it as a function of the source field, $\Omega(h)$, then performing a Legendre transformation,	
\begin{equation} \begin{aligned}
F(M) &= \Omega(h(M)) - h(M) M, \\
h(M) &= - \frac{\partial F(M)}{\partial M}.
\end{aligned} \end{equation}
This process is not trivial: $h(M)$ must be calculated by inverting $M(h) =  \partial \Omega(h) / \partial h$ of the reverse transform, which can be achieved by expanding both $h(M)$ and $M(h)$ to finite order in $U$ \cite{Fukuda1995}.
As a result of these expansions, a finite value of $M$ remains once the source field is set to zero at the end of the calculation -- required in order to retrieve the original theory -- and symmetry is spontaneously broken.

Variational perturbation theory may be viewed as a poor-man's method of source fields, that is much simpler to implement.
The variational terms that are added to and subtracted from the action can be identified with the source term, $S_{\text{source}} = h S_z$, and the term that is added as part of the Legendre transform, $- hM$, in the source fields method.
The key difference is that in the variational approach, the source field is assigned its leading order value, $h = -U\langle S_z \rangle$, from the outset \textit{via} the choice of variational terms, and the average is evaluated self-consistently in the presence of the order as $\langle S_z \rangle_{0,c}$ (see Eq.~\ref{eq:action_expansion}).

Collective order may alternatively be introduced into a system \textit{via} a Hubbard-Stratonovich transformation, which decouples the quartic electronic interaction in a desired channel to leave an effective theory for a collective field \cite{Conduit2009, Hannappel2016}.
Fluctuations in the collective field are then be evaluated in the presence of a finite, static ($i\omega_n=0$) component, which is the order parameter.
This differs from variational perturbation theory in which the order parameter is introduced variationally using a classical field, and  
fluctuations are evaluated using a diagrammatic perturbative treatment in terms of electronic fields.
A comprehensive comparison of these two methods is given in \cite{Kleinert2011}.
The advantage of the variational method is that multiple collective orders may be considered, in a democratic manner, simply by introducing variational terms for each.
This is (arguably) more straightforward than performing a Hubbard-Stratonovich decoupling of the electronic interaction in each channel.
The fluctuations are then evaluated about this multi-order state without risk of over-counting contributions \cite{Kleinert2011}.
However, the Hubbard-Stratonovich transformation benefits from being well-defined, without the ambiguity of having to define variational terms, and is perhaps preferable when an analysis of fluctuations in a particular channel (e.g. Cooper, nematic) is desired.

\section{\label{section:superconductivity}Two examples: fluctuation- driven superconductivity and spin nematicity}

In order to illustrate the methodology presented above, we will consider two examples: fluctuation-driven superconductivity and spin nematic order in the Hubbard model.
These are representative of orders in the particle-particle and particle-hole channels respectively.
Details of these calculations are presented in the Appendices.
	
\subsection{\label{subsection:QOBD_SC}Superconductivity}

We introduce spin-singlet superconductivity into the Hubbard model by adding and subtracting the variational terms,
\begin{equation} \begin{aligned}
S_{var}(\Delta) =  -\sum_k \Big( \Delta_{\bm k} c^\dagger_{k\uparrow} c^\dagger_{-k\downarrow} 
+ \bar{\Delta}_{\bm k} c_{-k\downarrow}c_{k\uparrow} \Big),
\end{aligned} \end{equation}
to the action, Eq. \ref{eq:Hubbard_action}.
These terms describes the conversion of two electrons into the superconducting condensate, described by
the order parameter, $\Delta_{\bm k}$, and its reverse process.
By studying $S_0^\prime$, the inverse propagator, $\mathcal G^{-1}(\Delta)$, of the ordered non-interacting system can be 	identified directly. In the basis of Nambu-Gorkov spinors, $\underline \psi_k^\dagger = (c^\dagger_{k\uparrow} \ c_{-k \downarrow} )$, this is $\mathcal G^{-1}_k (\Delta_{\bm k} ) = -i \omega_n \sigma_0 + \underline{\underline{h}}_{\bm k}$, where 	
$ \underline{\underline{h}}_{\bm k} = \epsilon_{\bm k} \sigma_z +  \left( \Delta_{\bm k} + \bar \Delta_{\bm k} \right) \sigma_x /2 + i  \left( \Delta_{\bm k} - \bar \Delta_{\bm k} \right) \sigma_y /2$ and $\sigma_i$ are Pauli matrices.
Inverting this matrix gives the Nambu-Gorkov propagator,
\begin{equation}  \begin{aligned}
\mathcal{G}_{\bm k}(i \omega_n) &= - \langle T \ \underline \psi_{\bm k} (i \omega_n) \otimes \underline \psi^\dagger_{\bm k} 
(i \omega_n) \rangle \\ 
&=  \begin{pmatrix}
G_{\bm k}(i \omega_n) & F_{\bm k} (i \omega_n) \\ F^\dagger_{\bm k}(i \omega_n) & -G^\dagger_{\bm k}(i \omega_n)
\label{eq:NG_GF}
\end{pmatrix},
\end{aligned} \end{equation}
where $G_{\bm k}(i \omega_n)$ and $F_{\bm k} (i \omega_n)$ are the normal and anomalous electronic propagators.

The expectation values in the free energy, Eq. \ref{eq:QOBD_free_energy}, are evaluated using these correlators.
A small value of the order parameter is assumed, and the result is expanded to $O(|\Delta|^2)$.
The most convenient way to achieve this is to expand the non-interacting propagators in powers of $\Delta$ before performing the Matsubara frequency summations.
 We find
\begin{equation} \begin{aligned}
G_{\bm k}(i\omega_n) &= G^0_{\bm k}(i\omega_n)  +  \bar{\Delta}_{\bm k} \ G^0_{\bm k}(i\omega_n) F^0_{\bm k}(i\omega_n)  + O(\Delta^4), \\
F_{\bm k}(i\omega_n) &= F^0_k(i\omega_n) +  O(\Delta^3), 
\label{eq:leading_order_propagators}
\end{aligned}\end{equation}
where $ F^0_k(i\omega_n) \equiv -\Delta_{\bm k} G^0_{\bm k}(i\omega_n) G^0_{\bm k}(-i\omega_n)$ and
$G^0_{{\bm k} \sigma}(i\omega_n) = (i\omega_n - \epsilon_{\bm k})^{-1}$ is the non-interacting propagator of the 
non-superconducting system.
We have once again absorbed $\mu$ into the dispersion.
Upon evaluating the Matsubara frequency sums, the free energy of the static ordered system is
\begin{equation} \begin{aligned}
F_{_{SO}}[\Delta]
&= F[\Delta=0]+  \sum_{\bm k} \ |\Delta_{\bm k}|^2 \ \frac{1-2n_{\bm k}}{2\epsilon_{\bm k}} \\
&+  U \sum_{{\bm k}{\bm k}^\prime}  \bar{\Delta}_{\bm k} \Delta_{{\bm k}^\prime} \
\frac{1-2n_{\bm k}}{2\epsilon_{\bm k}}\frac{1-2n_{{\bm k}^\prime}}{2\epsilon_{{\bm k}^\prime}},
\end{aligned}\end{equation}
where $n_{\bm k} \equiv n(\xi_{\bm k})$ and $\xi_{\bm k } = \sqrt{ \epsilon_{\bm k}^2 + |\Delta_{\bm k}|^2}$ is the Bogoliubov electronic dispersion of the superconducting state.
The second term is the anomalous Hartree term, which comes from evaluating $\langle S_{int} \rangle_{0,c}$.

When evaluating the fluctuation correction to the free energy we make two simplifying approximations. 
The first is that we assume that the frequency-symmetric part of the normal state electronic self-energy is negligible,
$\Sigma_{\bm k}(i\omega_n) \approx \left( \Sigma_{\bm k}(i\omega_n) - \  \Sigma_{\bm k}(-i\omega_n) \right)/2$.
This is an approximation often made in Eliashberg theory \cite{Eliashberg1960}.
It amounts to neglecting the terms in the free energy that are integrals over anti-symmetric functions of the electronic dispersion, $\epsilon_{\bm k}$, which are expected to vanish in the case of particle-hole symmetry.
The second is an on-shell approximation, which assumes that the electrons that form Cooper pairs occupy states at the Fermi level.
The way in which these approximations are applied is described in the Appendices.
We find the fluctuation-corrected free energy,
\begin{equation}\begin{aligned}
F[\Delta] &= F[\Delta=0] \\
&+ \sum_{\bm k}  |\Delta_{\bm k}|^2 \ \frac{1-2n_{\bm k}}{2\epsilon_{\bm k}}  
\left(1 - \frac{\partial \Sigma_{\bm k}(\epsilon_{\bm k})}{\partial \epsilon_{\bm k}} \Bigg|_{\epsilon_{\bm k}=0} \right) \\ 	
&+  \sum_{\bm k \bm q}  \bar \Delta_{\bm k} \Delta_{\bm k + \bm q} \ 
\frac{1- 2 n_{\bm k}}{2\epsilon_{\bm k}} \frac{1- 2 n_{\bm k + \bm q}}				
{2\epsilon_{\bm k + \bm q}} \ \Lambda_{\bm q}(0).
\label{eq:free_energy_SC1}
\end{aligned}\end{equation}
This is expressed in terms of the normal state self-energy,
\begin{equation} \begin{aligned}
&\Sigma_{\bm k}(i\omega_n) \\
&=  U^2 \sum_{\bm q} \sum_{i\omega_{n^\prime}} \ 
G^0_{\bm k + \bm q}(i\omega_{n^\prime}) \ \chi^0_{\bm q} (i\omega_n - i\omega_{n^\prime}) \\ 
&=  U^2 \sum_{\bm  p \bm q}
\frac{ n_{\bm p} (1 - n_{\bm p - \bm q})(1 - n_{\bm k + \bm q}) + (1-n_{\bm p})n_{\bm p - \bm q} n_{\bm k+ \bm q} }
{i\omega_n + \epsilon_{\bm p} - \epsilon_{\bm p - \bm q} - \epsilon_{\bm k + \bm q}}, 
\label{eq:SC_self_en}
\end{aligned}\end{equation}
in terms of the non-interacting susceptibility,
\begin{equation} \begin{aligned}
&\chi^0_{\bm q} (i\omega_n - i\omega_{n^\prime} ) \\
&= -  \sum_{\bm p} \sum_{i\omega_m} 
G^0_{\bm p \uparrow }(i\omega_m + i\omega_n - i\omega_{n^\prime}) 
G^0_{\bm p - \bm q \downarrow}(i\omega_{m})  \\
&= \sum_{\bm p} \frac{n_{\bm p} - n_{\bm p - \bm q } }
{i\omega_n - i\omega_{n^\prime}  + \epsilon_{\bm p - \bm q } - \epsilon_{\bm p }},
\label{eq:bare_suscept}
\end{aligned}\end{equation}
and the propagator of charge and spin fluctuations,
\begin{equation} \begin{aligned}
\Lambda_{\bm q}(i\omega_n - i\omega_{n^\prime}) 
= U +  U^2  \chi^0_{\bm q} (i\omega_n - i\omega_{n^\prime}).
\label{eq:SC_prop}
\end{aligned}\end{equation}
		
The second term in the free energy, Eq. \ref{eq:free_energy_SC1}, describes the renormalisation of the electronic quasiparticle 	states as a result of the superconducting order.
We can identify the (on-shell) field renormalisation, 
\begin{equation}
Z_{\bm k}(0) = 1 - \frac{\partial \Sigma_{\bm k}(\epsilon_{\bm k})}{\partial \epsilon_{\bm k}} \Bigg|_{\epsilon_{\bm k}=0},
\end{equation}
which takes Eliashberg form \cite{Eliashberg1960, Fay1980}. 
The bare vertex, $U$, does not contribute to the field renormalisation because the corresponding 
Hartree self-energy, $\Sigma^H_{\bm k}(i\omega_n) = U \sum_{\bm k^\prime} n_{\bm k^\prime}$, is symmetric in frequency 		and is therefore neglected.
The function $\Delta_{\bm k } (1-2n_{\bm k})/ (2\epsilon_{\bm k})  = \sum_{i\omega_n} F_{\bm k} ^0(i\omega_n)$ is interpreted as the (leading order) susceptibility of electrons of opposite momenta, $\{\bm k \uparrow, -\bm k \downarrow \}$, to forming a Cooper pair.
This function is sharply peaked about the chemical potential, which justifies the on-shell approximation.
The third term in Eq. \ref{eq:free_energy_SC1} drives the Cooper pairing of electrons, which is mediated by collective charge and spin fluctuations.
These are described in this theory by the non-interacting susceptibility, $\chi^0$, to which electrons couple with weight, $U$.
A well-established feature of fermionic condensation -- which is nevertheless worth mentioning here -- is that an effective interaction that is repulsive in the $s$-wave channel, $\Lambda>0$, may be attractive in higher angular momentum channels, in which it can couple areas of phase space across which the gap function changes sign \cite{Anderson1973}.

Spin-triplet superconductivity has previously been considered within the QOBD framework by Conduit \textit{et al.}  \cite{Conduit2013}.
In the case of A1-type pairing --  in which only electrons of same spin form Cooper pairs --  the result  
is very similar to Eq. \ref{eq:free_energy_SC1}, only with a few differences.
Firstly, pairing now occurs within a spin band, $\epsilon_{\bm  k\sigma}$, which gives a Bogoliubov dispersion $\xi_{\bm k \sigma} = \sqrt{\epsilon_{\bm  k\sigma}^2 + |\Delta_{\bm k}|^2}$, and so superconductivity is supported in a spin-polarised (e.g. ferromagnetic) system.
Secondly, the bare Hubbard interaction vertex has no contribution in the spin-triplet channel, i.e. the $O(U)$ free energy contribution is zero \citep{Berk1966, Fay1980}.
Instead, the fluctuation propagator is
$\Lambda_{\bm q \sigma}(i\omega_n - i\omega_{n^\prime}) 
=  U^2 \chi^0_{\bm q \sigma} (i\omega_n - i\omega_{n^\prime}) $, 
where the susceptibility is spin-dependent and described by the irreducible polarisation bubble,
$\chi^0_{q  \sigma} = - \sum_p G_{p \sigma} G_{p - q \sigma}$.
This result has the same form as the approximation to Eliashberg theory made by Fay and Appel \cite{Fay1980}.

\subsection*{Comparison to Eliashberg theory}

Any theory of superconductivity should be judged against Eliashberg theory \cite{Eliashberg1960}, which is the standard approach to treating superconducting systems in which electrons couple strongly to one another, \textit{via} some effective pairing interaction.
We will show that the QOBD theory of superconductivity of Eq.~\ref{eq:free_energy_SC1} may be viewed as a one-loop, linearised limit of Eliashberg theory, with a particular choice of pairing interaction.

Eliashberg theory is defined by the Dyson matrix equation in Nambu-Gorkov space,
\begin{equation}
\mathcal{G}^{-1}_k = \left( \mathcal{G}_{k}^{0} \right)^{-1} - \underline{\underline{\Sigma}}_k,
\label{eq:Eliashberg_G}
\end{equation}
where $\left( \mathcal{G}_{k}^{0} \right)^{-1} = \text{Diag} (i\omega_n - \epsilon_{\bm k}, \ i\omega_n + \epsilon_{\bm k})$, 
\begin{equation}
\underline{\underline{\Sigma}}_{\bm k}(i \omega_n) 
=  -\sum_{\bm k^\prime}\sum_{i \omega_{n^\prime}} \mathcal{G}_{\bm k^\prime} (i \omega_{n^\prime}) \ 	
\Lambda_{\bm k\bm k^\prime} (i \omega_{n} - i\omega_{n^\prime}), 
\label{eq:Eliashberg_Sigma}
\end{equation}
and $\Lambda$ is some effective electron-electron interaction.
This $GW$-like approximation to the Nambu-Gorkov self-energy neglects vertex corrections, which is justified in most cases of phonon-mediated interactions by Migdal's theorem \cite{Migdal1958,Chubukov2020}.
The self-energy matrix is also parameterised as 
\begin{equation} \begin{aligned}
\underline{\underline{\Sigma}}_{\bm k}(i \omega_n)
= \ &i \omega_n(1- Z_{\bm k}(i \omega_n)) \sigma_0 + \chi_{\bm k}(i \omega_n) \sigma_z \\ 
&+ \phi_{\bm k}(i \omega_n) \sigma_x -\bar{\phi}_{\bm k}(i \omega_n) \sigma_y  ,
\label{eq:Eliashberg_Sigma_param}
\end{aligned}\end{equation}
where $\sigma_i$ are Pauli matrices and both $Z_{\bm k}(i \omega_n)$ and $\chi_{\bm k}(i \omega_n)$ are even functions of frequency.
Equations \ref{eq:Eliashberg_G}, \ref{eq:Eliashberg_Sigma} and  \ref{eq:Eliashberg_Sigma_param} together give 			coupled self-consistency equations for the self-energy components which are the Eliashberg equations.
It is common to assume that the frequency-symmetric part of the normal self-energy is negligible,
$\chi_{\bm k}(i\omega_n) \approx 0$, and the anomalous self-energy is real, $\bar \phi_{\bm k}(i \omega_n) = 0$. 
This leaves two coupled equations,
\begin{equation} \begin{aligned}
i\omega_n(1- &Z_{\bm k}( i\omega_{n})) \\ 
&= \sum_{\bm k^\prime} \sum_{ i\omega_{n^\prime}} 
\frac{ i\omega_{n^\prime} Z_{\bm k^\prime}( i\omega_{n^\prime}) \ \Lambda_{\bm k\bm k^\prime}( i\omega_n - i\omega_{n^\prime})}{(i\omega_{n^\prime}  Z_{\bm k^\prime}( i\omega_{n^\prime}))^2 -
(\epsilon_{\bm 	k^\prime}^2 + \phi^2_{\bm k^\prime} ( i\omega_{n^\prime}))}, \\ 
\phi_{\bm k}( i\omega_{n}) &= - \sum_{\bm k^\prime} \sum_{ i\omega_{n^\prime}} 
\frac{ \phi_{\bm k^\prime}( i\omega_{n^\prime}) \ \Lambda_{\bm k \bm k^\prime}( i\omega_n - i\omega_{n^\prime}) }
{(i\omega_{n^\prime}  Z_{\bm k^\prime}( i\omega_{n^\prime}))^2 - (\epsilon_{\bm k^\prime}^2 + \phi^2_{\bm k^\prime}(i\omega_{n^\prime}))}.
\label{eq:Eliash_eqs}
\end{aligned} \end{equation} 
The superconducting gap function is identified as $\Delta_{\bm k}(i\omega_n) = \phi_{\bm k}(i\omega_n) /Z_{\bm k}(i\omega_n)$ from the poles of $\mathcal{G}_{k}$.
	
The QOBD result for the superconductor can be retrieved by taking Eliashberg theory in a series of approximations. 
Firstly, we assume that the system satisfies the classical (i.e. static) Eliashberg equations, i.e. 
$Z_{\bm k }(i \omega_n \neq 0) = 0$, $\phi_{\bm k }(i \omega_n \neq 0) = 0$.
Then the electronic excitations have a Bogoliubov-like dispersion,
$E_{\bm k}(i\omega_n = 0) = \sqrt{\tilde \epsilon_{\bm k}^2 + \Delta^2_{\bm k}}$,
where $\tilde \epsilon_{\bm k} \equiv \epsilon_{\bm k} / Z_{\bm k}(0)$ is the renormalised electronic dispersion and 
$\Delta_{\bm k} \equiv \Delta_{\bm k}(0) $ is the static component of the gap function, which is the superconducting order parameter. 
We also make a static approximation to the effective interaction, 
$\Lambda_{\bm k \bm k^\prime} (i \omega_n - i\omega_{n^\prime}) \approx \Lambda_{\bm k \bm k^\prime} (0)$.
Taken together, these approximations are equivalent to the on-shell approximation made in QOBD.

Secondly, we linearise the gap equation in $\Delta$;
\begin{equation} \begin{aligned}
\Delta_{\bm k} Z_{\bm k}(0) &= - \sum_{\bm k^\prime} \sum_{ i\omega_{n^\prime}} 
\Lambda_{\bm k \bm k^\prime}(0) \ 
\frac{ \Delta_{\bm k^\prime} Z^{-1}_{\bm k^\prime}( 0)}{(i\omega_{n^\prime} Z_{\bm k^\prime}(0) )^2  
\ + \ \tilde \epsilon_{\bm k^\prime}^2}.
\label{eq:Eliash_gap_eq}
\end{aligned} \end{equation} 
The on-shell field renormalisation, $Z_{\bm k}(0)$, may be written in terms of the energy derivative of the self-energy. 
Taking the gradient of the on-shell, frequency-antisymmetric self-energy,
$\Sigma^A_{\bm k} (\epsilon_{\bm k}) = \epsilon_{\bm k} (1- Z_{\bm k}(\epsilon_{\bm k}))$,
at the chemical potential, we find 
\begin{equation}
Z_{\bm k}(0)  = 1 - \frac{\partial \Sigma^{A}_{\bm k}(\epsilon_{\bm k})}{\partial \epsilon_{\bm k}} 
\Bigg|_{\epsilon_{\bm k} = 0}.
\end{equation}
Finally, we make a weak-coupling approximation.
To do so, we neglect terms that are higher that leading order in the interaction, $\Lambda$. 
This amounts to approximating $Z_{\bm k}(0) \approx 1$ on the right side of Eq.~\ref{eq:Eliash_gap_eq}.
In making this approximation we neglect the electron mass renormalisation, 
$\tilde \epsilon_{\bm k} \approx \epsilon_{\bm k}$.
It also means making a one-loop approximation to the normal state self-energy,
\begin{equation} \begin{aligned}
\Sigma_{\bm k}(i\omega_n) =  \sum_{\bm k^\prime}\sum_{i\omega_{n^\prime}}
G^0_{\bm k}(i\omega_{n^\prime}) \Lambda_{\bm k\bm k^\prime}( i\omega_n - i\omega_{n^\prime}).
\end{aligned} \end{equation} 
The Matsubara frequency sum in Eq. \ref{eq:Eliash_gap_eq} can now be evaluated straightforwardly to give
\begin{equation} \begin{aligned}
0 = \Delta_{\bm k} \left(1 - \frac{\partial \Sigma_{\bm k}(\epsilon_{\bm k})}{\partial \epsilon_{\bm k}} 
\Bigg|_{\epsilon_{\bm k} = 0} \right) 
+ \sum_{\bm k^\prime} \Delta_{\bm k^\prime} \frac{1-2n_{\bm k^\prime}}{2\epsilon_{\bm k^\prime}} 
\Lambda_{\bm k\bm k^\prime}(0).
\end{aligned} \end{equation} 
This is the QOBD gap equation -- the stationary equation of the free energy, Eq.~\ref{eq:free_energy_SC1} -- 
up to the choice of effective interaction, $\Lambda_{\bm k \bm k^\prime}$.
Setting $\Lambda_{\bm k \bm k^\prime}(0) = U + U^2 \chi^0_{\bm k \bm k^\prime}(0)$ retrieves the exact QOBD result for superconductivity in the Hubbard model.
In summary, the QOBD result for a superconductor can be thought of as a one-loop, $O(\Delta)$ limit of the zero-frequency Eliashberg equations.
				
\subsection{Spin nematic order}
\label{subsection:QOBD_nematic}

As an example of the QOBD approach to spin ordering, we now consider spin nematic order in the Hubbard model.
General spin order can be described by finite expectation values of the multipolar density operators in the spin-triplet 		channel.
In the momentum basis, these read 
\begin{equation} \begin{aligned}
\hat {\bm R}^t_{lm, \bm q} = \frac{1}{2} \sum_{\bm k } \underline c^\dagger _{\bm k + \bm q} \ \bm \sigma 
\ \Phi_{lm, \bm k} \ \underline c _{\bm k}, 
\label{eq:multipole}
\end{aligned}\end{equation}
where $l$ indexes the order of multipole expansion, $m$ indexes a component at that order (of which there are $2l+1$ possibilities) and $ \Phi_{lm, \bm k}$ is the corresponding form factor, which is a spherical harmonic in momentum space.
For example the $s$-wave channel, with form factor $\Phi_{l=0,m=0, \bm k} = 1$, gives the operator for a 					modulated spin density,  $\hat {\bm R}^t_{l=0,m=0, \bm q}  = \hat {\bm S}_{\bm q}$. 
The $\bm q = 0$ component describes a uniform spin density, whose expectation is the ferromagnetic moment.
	
We will consider spin nematic order, which corresponds to quadrupole ordering in the spin-triplet channel,
i.e. a finite $\langle \hat {\bm R}^t_{l=2,m, \bm q} \rangle$  \cite{Wu2007}.
The corresponding variational action that is used to introduce the order into the system describes the coupling of the spin-triplet, quadrupole density to a variational order parameter, $\eta$, which is its conjugate field.
In the absence of magnetic anisotropy or magnetic order, the three cartesian ordering directions are 	equivalent.
We choose ordering in the $z$ direction, which has a variational action,	
\begin{equation}
S_{var}(\eta) = \eta \sum_{k \sigma}  \sigma d_{\bm k} c^\dagger_{k\sigma} c_{k\sigma},
\label{eq:nematic_var}
\end{equation}
where $d_{\bm k} \equiv \Phi_{l=2,m, \bm k}$ is the form factor of a component in the $d$-wave channel.
In the uniform electron gas, spin nematicity amounts to spin-antisymmetric $d$-wave distortions of the spherical spin Fermi surfaces \cite{Hannappel2016, Karahasanovic2012}.
In this case the volumes of the spin Fermi surfaces are conserved and the order is non-magnetic.
	
The variational terms are introduced into the action of the Hubbard model, Eq.~\ref{eq:Hubbard_action}, before evaluating 	the effects of fluctuations about the ordered state up to $O(U^2)$ within the QOBD procedure.
The resulting free energy is 
\begin{equation} \begin{aligned}
F[\eta]= 
&-T \sum_{ \bm k\sigma} \ln \left(1+ e^{-\xi_{\bm k\sigma} /T} \right) +  \sum_{\bm k\sigma} \sigma d_{\bm k} n_{\bm k\sigma} \\ 
&+ U \sum _{\bm k \bm p} n_{\bm k\uparrow} n_{\bm p \downarrow} \\  
&+ U^2 \sum_{\bm k\bm p\bm q} 
\frac{ n_{\bm k+\bm q \uparrow} n_{\bm p-\bm q \downarrow} (1- n_{\bm p \downarrow})  (1- n_{\bm k \uparrow})}
{\xi_{\bm k+\bm q \uparrow} + \xi_{\bm  p - \bm q \downarrow} - \xi_{\bm p \downarrow} - \xi_{\bm k \uparrow}},
\end{aligned}\end{equation}	
where  $\xi_{\bm k \sigma}(\eta) = \epsilon_{\bm k} - \sigma d_{\bm k} \eta - \mu$ is the dispersion of the spin nematic state, and $n_{\bm k\sigma} \equiv n(\xi_{\bm k \sigma})$.
The effect of the order is therefore to deform the spin bands in a spin-antisymmetric way, \textit{via} the form factor $d_{\bm k }$.
This result was determined in \cite{Karahasanovic2012}.
We simplify it here by taking the same set of approximations as applied in the treatment of superconductivity above; namely, we assume  that the order parameter is small, that the self-energy is antisymmetric in frequency, and make an on-shell approximation.
This analysis is presented in the Appendices.
The free energy is then
\begin{equation} \begin{aligned}
F[\eta] &= F[\eta=0] \\
&- \eta^2 \sum_{\bm  k} d_{\bm k}^2 \ n^\prime (\epsilon_{\bm k})
\left(1 - 2 \frac{\partial \Sigma_{\bm k} (\epsilon_{\bm k}) }{\partial \epsilon_{\bm k}} \Bigg|_{\epsilon_{\bm  k} = 0} \right) \\
&- \eta^2 \sum_{\bm k\bm q} \ d_{\bm k} d_{\bm k + \bm q} \ n^\prime (\epsilon_{\bm k})n^\prime (\epsilon_{\bm k + \bm q}) \Lambda_{\bm q} (0),
\label{eq:nematic_expanded}
\end{aligned}\end{equation}
where $n^\prime (\epsilon_{\bm k}) = \partial n (\epsilon_{\bm k}) / \partial \epsilon_{\bm k}$ and the self-energy and 
fluctuation propagator are as defined in Eqs.~\ref{eq:SC_self_en} and \ref{eq:SC_prop} respectively.
Note that the factor of two in front of $\Sigma_{\bm k}$ is a spin factor.

This result has the same structure as the superconducting free energy, Eq. \ref{eq:free_energy_SC1}.
The second term accounts for the renormalisation of the electronic quasiparticle states due to the spin nematic order, described by the (on-shell) field renormalisation.
The third term drives ordering \textit{via} the fluctuation propagator.
This term appears with the opposite sign compared to the superconducting result.
As in the superconducting case, the effective interaction, $\Lambda$, may drive or disfavour ordering depending on the sign of the form factor, $d_{\bm k}$, in the areas of phase space being coupled.
The factor $-n^\prime (\epsilon_{ \bm k})$, which tends to $\delta(\epsilon_{\bm k})$ as $T\rightarrow 0$, 			counts the number of electronic states within the ordered band that coincide with the Fermi level (i.e. the density of states at the Fermi level is $D(\mu) = -\sum_{\bm k} n^\prime (\epsilon_{ \bm k})$). 
This behaves as the susceptibility to ordering in the particle-hole channel, which is counterpart to the anomalous susceptibility, $(1-2n(\epsilon_{\bm k}))/(2 \epsilon_{\bm k})$.

The $O(\eta^2)$ free energy has no minimum at finite values of $\eta$ and can only be used to find the temperature at which the system orders.
At $O(U)$, we find an ordering condition akin to the Stoner criterion, 
\begin{equation} \begin{aligned}
U \sum_{\bm k \bm k^\prime}  d_{\bm k} d_{\bm k^\prime}  n^\prime (\epsilon_{\bm k}) n^\prime 							(\epsilon_{\bm k^\prime}) \geq -\sum_{\bm k} d_{\bm k}^2  n^\prime (\epsilon_{ \bm k}) .
\end{aligned}\end{equation}		
Setting $d_{\bm k} =1$, such that $\eta =M$ describes ferromagnetic order along the $z$ direction, retrieves exactly the Stoner criterion, $U D(\mu) \geq 1$.
The simplest way to access terms higher order terms in $\eta$ is to allow the nematic order to modify the density of states, i.e. \textit{via} the replacement 
$n^\prime (\epsilon_{\bm k}) \rightarrow n^\prime (\xi_{\bm k \sigma})$, in Eq.~\ref{eq:nematic_expanded}, while paying attention to the spin structure of the result.
This approach will be taken later when considering spin nematic order in DFT.

\section{\label{section:QOBD-DFT}Introducing fluctuation-driven order into density functional theory}

Having illustrated the quantum order-by-disorder approach, we now use it to introduce fluctuation-driven electronic order into 	density functional theory.
This is achieved by applying QOBD, as presented above, to a system of Kohn-Sham electrons that are the solution of a spin-degenerate DFT calculation, which are now permitted to interact with one another.
In effect, the Kohn-Sham eigenstates provide a quasiparticle basis for the QOBD calculation.
Whereas a spin-degenerate DFT calculation is expected to reasonably capture charge order, it precludes spin order as well as order in the particle-particle channel, such as superconductivity.
In the framework presented here (which will be referred to as QOBD-DFT), non-charge order can be selectively introduced into DFT using quantum order-by-disorder, by means of a variational order parameter.
The advantage of this approach, compared to full spin- or superconducting DFT, is that ordering in specific channels may be considered in a controlled manner.
For example, in the case of spin ordering, the resulting optimisation of a single, scalar order parameter is much simpler 			that the full functional optimisation over the entire space of spin density matrices in spin-DFT.
		
The structure of the QOBD-DFT calculation is very similar to its application to the Hubbard model.
Order is introduced into the interacting Kohn-Sham system using variational terms in the action.
The free energy is then evaluated to second order in the interaction vertex, in terms of propagators of the static ordered system.
Now, we choose to calculate the spin and charge fluctuation propagators within DFT.
This renders the result fully first-principles, and also provides an improvement upon the leading order description of charge and spin fluctuations of QOBD.
Any double-counting of the effects of the electron-electron interaction is avoided by only keeping fluctuation corrections to the free energy of the Kohn-Sham system that vanish when the order parameter is taken to zero.
Minimising the free energy, which is a functional of the electron density, $n(\bm r)$, and the order parameter, gives 			coupled self-consistency equations, which are greatly simplified when the order is assumed to be a small perturbation 			on the Kohn-Sham system.
We will consider the same two examples as above: spin-singlet superconductivity and spin nematic order, and thus illustrate this approach for both particle-particle and particle-hole type orders.

\subsection{\label{subsection: KS_system}The interacting Kohn-Sham system}

First, the interacting Kohn-Sham system of electrons must be established.
We consider Kohn-Sham DFT at finite temperatures and at fixed chemical potential \cite{Pittalis2011, Lueders2005}, which we briefly review.
A finite-temperature Hohenberg-Kohn theorem \cite{Mermin1965} proves the existence of a one-to-one mapping between the electronic density function, $n(\bm r)$, and the external potential the system experiences, $v_{ext}( \bm r )$, in equilbrium.
As in the zero-temperature theory of Kohn and Sham \cite{Kohn1965}, the interacting system may be mapped onto an auxiliary (“Kohn-Sham") system of non-interacting electrons that has the same equilibrium electron density.
The thermodynamic potential of the auxiliary system may be written as a functional of the electron density,
\begin{multline}
\Omega^{\text{DFT}}[n(\bm r) ] = F_s[n(\bm r)]  + E_H[n(\bm r)] + F_{xc}[n(\bm r)] \\
+ \int d\bm r \ n(\bm r) \left( v_{ext}(\bm r) - \mu \right).
\label{eq:DFT_free_energy}
\end{multline}
$F_s = T_s - T S_s$ contains contributions from the kinetic energy and the entropy of the Kohn-Sham electrons, $E_H$ is the Hartree energy and $F_{xc}$ is the free energy of exchange and correlations. 
The latter is commonly approximated using its ground state value, $E_{xc}[n(\bm r)]$ \cite{Pittalis2011, Lueders2005}, which is an approximation that we also make here.
$\Omega^{\text{DFT}}[n(\bm r)]$ is minimised by the equilibrium electron density, at which it takes the value of the grand potential of the interacting system in question.

The electron density is expressed in terms of the single-electron eigenstates of the auxiliary system, 
\begin{equation}
n(\bm r)= \sum_{n \bm k} n(\epsilon_{n\bm k} - \mu) |\phi_{n \bm k}(\bm r)|^2,
\label{eq:electron_density}
\end{equation}
which are thermally occupied. 
These eigenfunctions have been expressed as Bloch wavefunctions, $\phi_{n\bm k}(\bm r)=e^{i\bm k \cdot \bm r}u_{n \bm k}(\bm r)$, with band, $n$, and crystal momentum, $\bm k$, indices.
Minimising $\Omega^{\text{DFT}}[n(\bm r)]$  with respect to $n(\bm r)$ gives the $T=0$ Kohn-Sham equations,
\begin{equation}
\Big( -\frac{\nabla^2}{2}  + v_{_{\text{KS}}}(\bm r) - \mu \Big) \phi_{n \bm k}(\bm r) = \epsilon_{n \bm k} \ \phi_{n \bm k}(\bm r) .
\label{eq:KS_equations}
\end{equation}
The Kohn-Sham potential,
\begin{equation}
v_{_{\text{KS}}}(\bm r) =  \frac{\delta E_H[n(\bm r)]}{\delta n(\bm r)} + \frac{\delta E_{xc}[n(\bm r)]}{\delta n(\bm r)} \ + \ v_{ext}(\bm r ),
\end{equation}
is the effective potential that the Kohn-Sham electrons experience such that they reproduce the equilibrium electron density of the 	interacting system.
Therefore the only effect of the finite temperature is to modify the electron density as in Eq.~\ref{eq:electron_density}.
	
When applying quantum order-by-disorder to DFT, we consider a system of \textit{interacting} Kohn-Sham electrons.
These are electrons that occupy the eigenstates of the Kohn-Sham auxiliary system, which we allow to interact \textit{via} an instantaneous, screened Coulomb interaction between opposite spins, $V(| \bm r- \bm r^\prime|)$.
For now, we will leave the form of this interaction vertex unspecified.
The reason for this is that, in the end, the charge and spin fluctuation propagators will be calculated from first-principles, using results from time-dependent DFT (reviewed in Section \ref{subsection:DFT_fluct_prop}), in which the screened interaction vertex is described by the exchange-correlation kernel. 
Our approach is therefore to determine the structure of the QOBD-DFT theory for arbitrary $V(| \bm r- \bm r^\prime|)$, before retrofitting the result with the DFT spin and charge fluctuation propagators. The result is a fully first-principles theory.

The starting point of the QOBD-DFT calculation is the Hamiltonian, 
\begin{equation}\begin{aligned}
\hat H = \sum_\sigma &\int d \bm r \ c^\dagger_\sigma (\bm r)
\left( -\frac{\nabla^2}{2} + v_{_{\text{KS}}}(\bm r) - \mu \right) c_\sigma(\bm r) \\
+ &\int d \bm r d \bm r^\prime \ \hat n_\uparrow(\bm r)V(|\bm r- \bm r^\prime|) \hat n_\downarrow(\bm r^\prime),
\label{eq:interacting_KS_H}
\end{aligned}\end{equation}
where $c^\dagger_\sigma (\bm r) $ creates an electron with spin $\sigma$ at position $\bm r$.
Transforming to the basis of Kohn-Sham eigenstates, 
$c^\dagger_\sigma (\bm r)=\sum_{n \bm k} \phi^{*} _{n \bm k}(\bm r) c^\dagger_{{n \bm k}\sigma}$, gives
\begin{equation} \begin{aligned}
&\hat H=\sum_{n \bm k \sigma} \left(\epsilon_{n \bm k} - \mu \right) c^\dagger_{n \bm k\sigma}c_{n \bm k \sigma} \\
&+ \sum_{(n\bm k)_{1,2,3,4}} V_{n_1\bm k_1, n_2 \bm k_2, n_3 \bm k_3 ,n_4 \bm k_4} 
c^\dagger_{n_1\bm k_1 \uparrow}  c^\dagger_{n_2 \bm k_2\downarrow} c_{n_3 \bm k_3 \downarrow} c_{n_4 \bm k_4 \uparrow}.
\label{eq:interacting_KS}
\end{aligned}\end{equation}
The matrix elements of the interaction vertex are
\begin{multline}
V_{n_1\bm k_1, n_2 \bm k_2, n_3 \bm k_3 ,n_4 \bm k_4} 
= \delta_{\bm k_1 - \bm k_4, \ \bm k_3 - \bm k_2}  \int_{V_{cell}} d \bm r d \bm r^\prime \\
\phi^*_{n_1\bm k_1}(\bm r) \phi^*_{n_2 \bm k_2}(\bm r^\prime)  V(|\bm r- \bm r^\prime|) \phi_{n_3 \bm k_3}(\bm r^\prime) \phi_{n_4 \bm k_4} (\bm r),
\end{multline}
where the $\delta$-function conserves crystal momentum.
This Hamiltonian is the starting point of the QOBD calculation.
It then proceeds exactly as described above: 
variational terms that describe the coupling of the Kohn-Sham electrons to a desired order parameter are introduced into the 		action of the interacting Kohn-Sham system, as in Eq. \ref{eq:variational_action}.		
The correlators of the ordered non-interacting system are identified, and the fluctuation-corrected free energy, Eq.~\ref{eq:QOBD_free_energy}, is evaluated in this basis.

\subsection{\label{subsection:SC_DFT}Superconductivity}
	
Here we apply this approach to introduce spin-singlet superconductivity into DFT.
Unlike in the single-band Hubbard model studied above, we now have many bands to deal with, which means that care must be taken when defining the variational terms.
For example, in the case of superconductivity, there may be multiple electron bands in the vicinity of the Fermi level 	that support Cooper pairing, and pairing may occur intra- or inter-band.
The variational terms that allow all such possibilities are
\begin{equation}
S_{var}[\Delta] =  -\sum_{nn^\prime} \sum_{k} \left(\Delta_{nn^\prime \bm k} \ 
c^\dagger_{n  k \uparrow} c^\dagger_{n^\prime -  k \downarrow}  + \text{h.c.} \right).
\label{eq:SC_var_action}
\end{equation}
$\Delta_{nn^\prime \bm k}$ is the gap function of the superconducting Kohn-Sham system, which is strictly defined as the matrix elements of a pairing potential.
As shown in a formally-derived superconducting DFT \cite{Lueders2005}, this is an anomalous counterpart to the normal Kohn-Sham potential that ensures that the Cooper pair densities of the Kohn-Sham and interacting systems are equal.
Like $v_{_{\text{KS}}}$, this is a property of the auxiliary non-interacting system and is therefore not strictly equal to the true gap function of the interacting system \cite{Marques2005, Sanna2020}. 
Nevertheless, $\Delta_{nn^\prime \bm k}$ is expected to be a reasonable approximation to its true value, as is generally the case for the Kohn-Sham electronic bands in the normal state.
For simplicity we will assume that ordering occurs within a single band, with index $\alpha$, that exists at the Fermi level, i.e. 
$\Delta_{nn^\prime \bm k} = \Delta_{nn^\prime \bm k}  \delta_{n \alpha}  \delta_{n^\prime \alpha}$.
	
We omit the details of the calculation and the approximations employed, as these are the same as in the application to the Hubbard model above.
The free energy of the superconducting Kohn-Sham system has the same form,
\begin{equation} \begin{aligned}
&F[\Delta] =  F[\Delta=0]  \\
&+ \sum_{\bm k}  |\Delta_{\alpha \bm k}|^2 \ \frac{1-2n (\epsilon_{\alpha \bm  k})} {2\epsilon_{\alpha \bm  k}}  
\left(1 - \frac{\partial \Sigma_{\alpha \bm  k}(\epsilon_{\alpha \bm  k})}{\partial \epsilon_{\alpha \bm  k}} 
\Bigg|_{\epsilon_{\alpha \bm k}=0} \right) \\	
&+ \sum_{\bm k\bm k^\prime} \ \bar \Delta_{\alpha \bm k} \Delta_{\alpha \bm k^\prime} \ 
\frac{1- 2 n(\epsilon_{\alpha \bm k}) }{2\epsilon_{\alpha \bm k}} \frac{1- 2 n(\epsilon_{\alpha \bm k^\prime})}		
{2\epsilon_{\alpha \bm k^\prime}} \Lambda_{\alpha \bm k, \alpha \bm k^\prime}(0),
\label{eq:QOBD_SC_free_energy}
\end{aligned}\end{equation}
where the chemical potential, $\mu$, has been absorbed into the Kohn-Sham dispersions, $\epsilon_{n\bm k}$.
The field renormalisation term depends on the one-electron self-energy of the superconducting band, where 
\begin{multline}
\Sigma_{n \bm k}(i\omega_n) =    \sum_{n^\prime m m ^\prime} \sum_{\bm k^\prime \bm p \bm p^\prime} \sum_{i\omega_{n^\prime}} \ 
|V_{n\bm k, n^\prime \bm k^\prime, m \bm p, m^\prime \bm p^\prime}|^2 \\
G^0_{n^\prime \bm k^\prime}(i\omega_{n^\prime}) \ 
\chi^0 _{m\bm p, m^\prime \bm p^\prime}( i\omega_n - i\omega_{n^\prime}),
\label{eq:KS_SF_propagator}
\end{multline}
and $G^0_{n \bm k}(i\omega_{n}) = \left(i\omega_{n} - \epsilon_{n \bm k}\right)^{-1}$.
The matrix elements of the fluctuation propagator are 
\begin{equation} \begin{aligned}
&\Lambda_{n \bm k, n^\prime \bm k^\prime}( i\Omega_m) =  
V_{n\bm k, n^\prime \bm k^\prime, n^\prime \bm k^\prime, n\bm k} \\
&+  \sum_{mm^\prime} \sum_{ \bm p \bm p^\prime} V_{n\bm k, n^\prime \bm k^\prime, m \bm p, m^\prime \bm p^\prime} 
\chi^0 _{m \bm p, m^\prime \bm p^\prime}( i\Omega_m) V_{m^\prime \bm p^\prime, m \bm p, n^\prime \bm k^\prime, n \bm k},
\end{aligned}\end{equation}
in terms of the matrix elements of the non-interacting susceptibility,
\begin{equation} \begin{aligned}
\chi^0 _{m \bm p, m^\prime \bm p^\prime}( i\Omega_m) =
\frac{n(\epsilon_{m \bm p})- n(\epsilon_{m^\prime \bm p^\prime}) }{i \Omega_m + \epsilon_{m^\prime \bm p^\prime} - \epsilon_{m \bm p}},
\label{eq:KS_NI_susceptibility}
\end{aligned}\end{equation}
where $\Omega_m = 2m \pi T$ is the bosonic Matsubara frequency that is carried by an electron-hole pair.
Note that the conservation of crystal momentum is enforced throughout by $\delta$-function in the interaction vertex, $V$.
This result is effectively a generalisation of the QOBD result for the single-band Hubbard model to a multi-band system of Kohn-Sham electrons.
Whereas superconducting has only been permitted in one band, the susceptibility (i.e. the effects of charge and spin fluctuations) must be evaluated while taking into account electron-hole excitations across all occupied and unoccupied bands.

\subsection{\label{subsection:DFT_fluct_prop}Calculating fluctuation propagators within DFT}

The key quantities that enter this theory are the charge and spin susceptibilities, which take their leading order values,  $\chi^0$,  and the vertex, $V$, that couples electrons to collective fluctuations.
The interaction vertex has not yet been specified, and instead of considering some phenomenological, screened electronic interaction \cite{vonKeyserlingk2013} we choose to calculate the full spin and charge fluctuation propagators within DFT.
These are well-established results from time-dependent DFT \cite{Northrup1989, Cao2018}; we review them here for the sake of a self-contained theory.
This approach has two advantages.
Firstly, it renders the QOBD-DFT calculation fully first-principles, and secondly, it improves upon the approximations of QOBD.
The DFT charge and spin susceptibilities are non-perturbative results that exceed the leading order ($O(V^2)$) approximation to fluctuations made in QOBD. 
Indeed, the DFT interaction vertex in principle captures vertex corrections to the self-energy that are not even included in Eliashberg theory \cite{Romaniello2012}.
				
In a non-magnetic, non-spin-orbit-coupled system, the charge, $\Lambda^C$, and spin, $\Lambda^S$, fluctuation propagators of Kohn-Sham DFT can be written as
\begin{widetext}
\begin{equation} \begin{aligned}
\Lambda^C(\bm r, \bm r^\prime; i\Omega_m)  
&= v_c(\bm r - \bm r^\prime) + f^C_{xc}(\bm r, \bm r^\prime) +  \int d \bm r_1 d\bm r_2 \ \left[ v_c(\bm r - \bm r_1 )  + f^C_{xc}(\bm r, \bm r_1) \right] \chi^C (\bm r_1, \bm r_2; i\Omega_m) \left[ v_c(\bm r_2 - \bm r^\prime) + f^C_{xc}(\bm r_2, \bm r^\prime) \right], \\
\Lambda^S(\bm r, \bm r^\prime; i\Omega_m) 
&= 3 \left( f_{xc}^S(\bm r, \bm r^\prime) + 
\int d \bm r_1 d\bm r_2 \ f_{xc}^S(\bm r, \bm r_1) \ \chi^S (\bm r_1, \bm r_2; i\Omega_m) \ f_{xc}^S(\bm r_2, \bm r^\prime) \right),
\label{eq:DFT_fluct_propagators}
\end{aligned} \end{equation}
\end{widetext}
in terms of the charge and spin susceptibilities, 
\begin{equation} \begin{aligned}
\chi^C(\bm r, \tau; \bm r^\prime, \tau^\prime) &=\frac{\delta n(\bm r,\tau) }{ \delta E(\bm r^\prime,\tau^\prime)}, \\
\chi^S(\bm r, \tau; \bm r^\prime, \tau^\prime) &= \frac{\delta m(\bm r,\tau) }{ \delta B(\bm r^\prime,\tau^\prime)}.
\label{eq:DFT_susceptibilities}
\end{aligned}\end{equation} 
$n$ and $m$ are charge and magnetisation densities respectively, $E$ and $B$ are perturbing electric and magnetic fields, and $\tau$ is an imaginary time variable.
$v_c(\bm r - \bm r^\prime)$ is the bare Coulomb interaction, which only contributes in the charge channel.
The effective interaction vertices in the charge and spin channels are the corresponding exchange-correlation kernels,
\begin{equation} \begin{aligned}
f_{xc}^C(\bm r, \bm r^\prime) = \frac{\delta v_{xc}(\bm r) }{\delta n(\bm r^\prime)}, 
\quad f_{xc}^S(\bm r, \bm r^\prime) = \frac{\delta v_{xc}(\bm r) }{\delta m(\bm r^\prime)},
\label{eq:DFT_kernels}
\end{aligned} \end{equation} 
where $v_{xc}[n(\bm r),m(\bm r)](\bm r)$ is the Kohn-Sham potential of spin-DFT.
Note that the factor of 3 in $\Lambda^S$ is a result of equal contributions of spin fluctuations in the three spatial directions. 	

As shown by Essenberger et al. \cite{Essenberger2014}, these results follow from a local approximation to the electronic self-energy using the spin-resolved exchange-correlation potential, 
$\Sigma_\sigma(\bm r, \tau; \bm  r^\prime, \tau^\prime) \approx v_{xc \ \sigma}(\bm  r, \tau) \ \delta(\bm r-\bm r^\prime) \delta(\tau-\tau^\prime)$.
The irreducible particle-hole propagator, which is a 4-point function of position and imaginary time, is then calculated as 
$\Lambda^0_{\sigma \sigma^\prime} = \delta \Sigma_\sigma / \delta G_{\sigma^\prime}$, and the full particle-hole propagator is found by solving the Bethe-Salpeter equation, $\Lambda = \Lambda^0 + \Lambda^0 GG \Lambda$.
Equivalently, these results can be derived within time-dependent spin density functional theory, within a linear 				response framework	\cite{Gross1985,Cao2018}.
		
An adiabatic approximation has been applied in Eq.~\ref{eq:DFT_fluct_propagators}, as is common in time-dependent DFT \cite{Gross1990, Cao2018}.
The system is assumed to always remain in its ground state, and only the ground state densities vary with time, i.e. 
$v_{xc}[n(\bm r,\tau), m(\bm r,\tau) ](\bm r,\tau) \approx v_{xc}[n(\bm r,\tau), m(\bm r,\tau)](\bm r)$.
Then the vertices are local in time,
\begin{equation}\begin{aligned}
f^{C}_{xc}(\bm r, \tau; \bm r^\prime, \tau^\prime) 
=  \frac{\delta v_{xc}(\bm r,\tau) }{\delta n(\bm r^\prime,\tau^\prime)} 
=f^{C}_{xc}(\bm r, \bm r^\prime) \delta(\tau - \tau^\prime),
\end{aligned}\end{equation}
and similarly for $f^S_{xc}$. All frequency dependence is then carried solely by the susceptibilities.
Assuming a local spin density approximation to $v_{xc}(\bm r)$, the exchange-correlation kernels are also local in space, 
\begin{equation}
f^C_{xc}(\bm r, \bm r^\prime) 
= \frac{\delta v^{^{\text{LSDA}}}_{xc}[n(\bm r ), m(\bm r )](\bm r)}{ \delta n(\bm r)} \delta(\bm r -\bm r^\prime).
\label{eq:xc_kernels}
\end{equation}
The kernel $f^C_{xc}$ contains local vertex corrections \cite{Romaniello2012}.
This can be seen by setting $f^C_{xc}=0$ in Eq.~\ref{eq:DFT_fluct_propagators}, which retrieves the effective interaction, 		$W$, of the $GW$ approximation, in which vertex corrections are absent by definition.
		
The charge and spin susceptibilities of Eq. \ref{eq:DFT_susceptibilities} may be calculated by solving their Dyson matrix equations
\cite{Cao2018, Staunton1999, Lischner2014, Lischner2015}, written schematically as
\begin{equation} \begin{aligned}
\chi^C &= \chi^0 + \chi^0 \left(v_c + f^C_{xc}\right)\chi^C, \\
\chi^S &= \chi^0 + \chi^0 f^S_{xc} \ \chi^S,
\label{eq:sus_Dyson}
\end{aligned} \end{equation}
in terms of the bare susceptibility, Eq.~\ref{eq:KS_NI_susceptibility}.
Alternatively, the susceptibilities may be found by solving the Sternheimer equations of time-dependent density functional perturbation theory 	\cite{Savrasov1998}, which are equations for the leading corrections to the Kohn-Sham wavefunctions due to a time-dependent perturbation. 		
In this case, the susceptibilities are evaluated as derivatives as in Eq. \ref{eq:DFT_susceptibilities}. 
The advantage of the Dyson equation approach is that, when considering paramagnetic systems in the absence of spin-orbit 		coupling, the spin and charge fluctuation propagators can both be calculated within \textit{spin-degenerate} DFT.
The LSDA exchange-correlation kernels, $f_{xc}^{S,C}$, are calculated analytically at $m(\bm r) = 0$ \cite{Lischner2015}.
	
\subsection*{Retrofitting the QOBD-DFT result} 

The QOBD-DFT results are retrofitted with the DFT fluctuation propagators simply by replacing the fluctuation propagators, 
$\Lambda$, in the free energy with their DFT counterparts, Eq. \ref{eq:DFT_fluct_propagators}.
In the case of spin-singlet superconducting order (Eq. \ref{eq:QOBD_SC_free_energy}) the QOBD-DFT free energy then reads 
\begin{widetext}
\begin{equation} \begin{aligned}
F[\Delta] = \ &F[\Delta=0] \ + \ \sum_{\bm k}  |\Delta_{\alpha \bm k}|^2 \ \frac{1-2n (\epsilon_{\alpha \bm k})} {2\epsilon_{\alpha \bm k}}  
\left(1 - \frac{\partial \Sigma_{\alpha \bm k}(\epsilon_{\alpha \bm k})}{\partial \epsilon_{\alpha \bm k}} 
\Bigg|_{\epsilon_{\alpha \bm k}=0} \right) \\ 	
&+ \ \sum_{\bm k \bm k^\prime} \ \bar \Delta_{\alpha \bm k} \Delta_{\alpha \bm k^\prime} \ 
\frac{1- 2 n(\epsilon_{\alpha \bm k}) }{2\epsilon_{\alpha \bm k}} \frac{1- 2 n(\epsilon_{\alpha \bm k^\prime})}
{2\epsilon_{\alpha \bm k^\prime}}  
\Big( \Lambda^C_{P \ \alpha \bm k, \alpha \bm k^\prime}(0) + \Lambda^S_{P \ \alpha \bm k, \alpha \bm k^\prime}(0) \Big),
\label{eq:QOBD_SC_retro}
\end{aligned}\end{equation}
where
\begin{equation} \begin{aligned}
\Sigma_{n \bm k}(i\omega_n) 
&=   \sum_{n^\prime \bm k^\prime } \sum_{i \omega_{n^\prime}} G^0_{n^\prime \bm k^\prime}(i\omega_{n^\prime}) 
 \Lambda^S_{R \ n \bm k, n^\prime \bm k^\prime} (i\omega_n - i\omega_{n^\prime}), \\ 
\Lambda^{C,S}_{P \ n \bm k, n^\prime \bm k^\prime}(i\Omega_m) 
&= \int d \bm r d\bm r^\prime \phi^*_{n \bm k}(\bm r) \phi^*_{n -\bm k}(\bm r^\prime) \ 
 \Lambda^{C,S}(\bm r, \bm r^\prime; i\Omega_m) \ 
\phi_{n^\prime -\bm k^\prime}(\bm r^\prime) \phi_{n^\prime \bm k^\prime}(\bm r), \\ 
\Lambda^S_{R \ n \bm k, n^\prime \bm k^\prime} (i\Omega_m) 
&= \int d\bm r d \bm r^\prime \phi^*_{n^\prime \bm k^\prime}(\bm r)  \phi^*_{n\bm k}(\bm r^\prime) 
 \  \Lambda^S(\bm r, \bm r^\prime; i\Omega_m)\ \phi_{n^\prime \bm k^\prime}(\bm r^\prime) \phi_{n \bm k}(\bm r).
\label{eq:QOBD_SE_SF_propagators}
 \end{aligned}\end{equation}
\end{widetext}
This is the free energy functional for a superconducting system.
The $\Delta$-dependent terms are a superconducting correction to the exchange-correlation functional.
It is not an explicit functional of the electron density, but rather an orbital-dependent density functional \cite{Kummel2008}.
$\Lambda^{C,S}_{n \bm k, n^\prime \bm k^\prime}(i\Omega_m)$ are the matrix elements of the DFT fluctuation propagators in the basis of Kohn-Sham orbitals.
These are defined differently in the field renormalisation ($\Lambda_R$) and pairing ($\Lambda_P$) terms.
In the pairing term, $\Lambda$ scatters a Cooper pair $(n \bm k \uparrow, n$-$ \bm k\downarrow)$ to  
$(n^\prime \bm k^\prime \uparrow, n^\prime$-$\bm k^\prime\downarrow)$, whereas in the field renormalisation term, it 			scatters an electron in a state $(n\bm k \sigma)$ into a state $(n^\prime \bm  k^\prime \sigma)$.
Moreover, the effects of charge fluctuations, $\Lambda^C$, do not appear in the field renormalisation term.
It is assumed that all the effects of charge fluctuations in the normal state that are captured by the local density approximation to $\Lambda^C$ are already accounted for in the normal KS-DFT calculation \cite{Essenberger2014}, which enter the 			calculation \textit{via} the dispersions, $\epsilon_{n \bm k}$.

This result leaves a Matsubara frequency sum to be evaluated in the self-energy, in Eq.~\ref{eq:QOBD_SE_SF_propagators}, to which there are two possible approaches.
The first is to perform this sum numerically; the second is to assume that $\Lambda^{C,S}_{n \bm k, n^\prime \bm k^\prime}( i\Omega_m)$ have simple pole structures, i.e. a Kramers-Kronig relation,	
\begin{equation}
\Lambda_{n \bm k, n^\prime \bm k^\prime} (i\Omega_m) 
= \int \frac{d \omega}{\pi} \frac{1}{\omega - i\Omega_m} \text{Im}[\Lambda_{n \bm k, n^\prime \bm k^\prime}  (\omega)],
\end{equation}
in terms of the spectral function, $\text{Im}[\Lambda_{\bm k \bm k^\prime} (\omega)]$.
Then the Matsubara frequency sum can once again be performed analytically, in exchange for a real-axis integral over  		frequencies, $\omega$ \cite{Essenberger2014}.	

\subsection*{Comparison to superconducting DFT} 

A superconducting density functional theory (SCDFT) that supports fluctuation- (as well as phonon-) driven pairing has previously been developed \cite{Lueders2005, Marques2005, Essenberger2014}.
The approach of SCDFT differs from the QOBD-DFT method in that it is formulated explicitly in terms of functionals of a Cooper pair density function.
This is achieved in a formally exact way by means of a Hohenberg-Kohn theorem for the pair density \cite{Oliveira1988}.
The non-interacting Kohn-Sham system is then constrained to have electron and Cooper pair densities that are equal to those of the interacting system, which is ensured by their conjugate potentials: the normal and anomalous (or ‘pairing') Kohn-Sham potentials.

This approach differs from that of QOBD-DFT, which is formulated directly in terms of the superconducting gap function -- matrix elements of the Kohn-Sham pairing potential -- that is introduced as a variational parameter.
Yet in spite of this difference in approaches, the results are the same (up to the additional, on-shell approximation made in this work).
This is evident when comparing the free energy, Eq. \ref{eq:QOBD_SC_retro}, to the superconducting gap equation in \cite{Essenberger2014}.
This is no surprise since the pair density and the pairing potential are conjugate variables, and thus the theory can equivalently be formulated in terms of either quantity.
Moreover, the treatment of charge and spin fluctuations in both theories are equivalent, involving a one-loop evaluation of the self-energy in terms of the DFT fluctuation propagators, Eq. \ref{eq:QOBD_SE_SF_propagators}.
The ‘decoupling approximation' that is made in SCDFT is effectively enforced in QOBD-DFT \textit{via} the choice of variational terms, Eq.~\ref{eq:SC_var_action}, which insists that the gap function is diagonal in the crystal momentum basis.
						
\subsection{\label{subsection:SC_implement}An implementation scheme for superconducting order}

Determination of the equilibrium state of the superconducting Kohn-Sham system requires that its free energy (or strictly, its thermodynamic potential) is minimised with respect to electron density (or equivalently the Kohn-Sham orbitals, $\phi_{n\bm k}$) and the gap function, $\Delta_{\bm k}$.
Their minimising values are described by stationary equations, which will now be determined in a few simplifying approximations.
Taking the free energy, Eq.~\ref{eq:QOBD_SC_retro}, we first identify the non-superconducting part, $F[\Delta=0]$, with the 
thermodynamic potential of Kohn-Sham DFT, $\Omega^{\text{DFT}}$, in Eq. \ref{eq:DFT_free_energy}.
Labelling the order-dependent part as $F^{\Delta}$, the resulting thermodynamic potential is 
\begin{equation} \begin{aligned}
&\Omega[n(\bm r), \{ \epsilon_{n \bm k}, \phi_{n \bm k}, \Delta_{n \bm k}\}]  \\
&= \Omega^{DFT}[n(\bm r), \{\phi_{n \bm k} \} ] 
+  F^{\Delta}[\{ \epsilon_{n \bm k}, \phi_{n \bm k}, \Delta_{n \bm k}\}],
\label{eq:free_energy_minimise}
\end{aligned} \end{equation}
which is to be minimised.
$F^{\Delta}$ can be interpreted as the correction to the normal 	exchange-correlation energy functional, $ F_{xc}$, due to the additional electronic correlations that are present in the superconducting phase. 
Since $F^\Delta$ vanishes when $\Delta =0$, any double-counting of exchange-correlation effects is prevented.
		
When minimising the thermodynamic potential it is important to take into account the dependence of the electron density, $n(\bm r)$, on both $\phi_{n \bm k}( \bm r)$ and  $\Delta_{n \bm k}$. 
This is because the electron density of the Kohn-Sham system is renormalised by the superconducting order,
 \begin{equation} \begin{aligned}
n( \bm r)  
&= \sum_{n \bm k} \langle c_{n \bm k}^\dagger( \bm r)  c_{n \bm k} ( \bm r)  \rangle \\ 
&= \sum_{n \bm k}  \left(1 - \frac{\epsilon_{n \bm k} }{\xi_{n \bm k} } (1 - 2n(\xi_{n \bm k} )) \right) |\phi_{n \bm k} ( \bm r)|^2,
\label{eq:renormalised_density}
\end{aligned} \end{equation}
where a Bogoliubov transformation, $\alpha_{{n \bm k} \sigma}^\dagger = u_{n \bm k} c^\dagger_{{n \bm k} \sigma} + \sigma \bar{v}_{n \bm k} c_{n -\bm k-\sigma}$, has been applied (see Appendix \ref{Appendix:SC}), and
$\xi_{n \bm k} = \sqrt{ \epsilon_{n \bm k} ^2 + |\Delta_{n \bm k}|^2}$ is the Bogoliubov dispersion of the Kohn-Sham system.	
When superconductivity is weak, the variation of the electron density with respect to the Kohn-Sham orbitals is
\begin{equation}\begin{aligned} 
\frac{\delta n( \bm r)}{\delta \phi_{n \bm k}^*( \bm r)} &= \sum_{\sigma} n(\epsilon_{n\bm k}) \phi_{n\bm k} ( \bm r) \\
&+ \sum_\sigma \frac{|\Delta_{n \bm k}|^2}{2\epsilon_{n \bm k}} \left( \frac{1-2n(\epsilon_{n \bm k})}{2\epsilon_{n \bm k}} 
+  \frac{\partial n(\epsilon_{n \bm k})}{\partial \epsilon_{n \bm k}} \right) \phi_{n\bm k} ( \bm r) \\
&+ \ O(|\Delta|^4). 
\end{aligned}\end{equation}
This result is simplified by noting that, when the superconducting bands are sufficiently close to the Fermi level, the 			superconducting susceptibility function becomes an approximation to the energy derivative of the Fermi function, 	
\begin{equation} \begin{aligned}
&\lim_{\epsilon_{n \bm k } \to 0} \frac{1-2n(\epsilon_{n \bm k})}{2\epsilon_{n \bm k}} \\
&\quad \quad= \lim_{\epsilon_{n \bm k } \to 0} - \frac{n(\epsilon_{n \bm k}) -  n(-\epsilon_{n \bm k})}{2\epsilon_{n \bm k}}  
\approx - \frac{\partial n(\epsilon_{n \bm k})}{\partial \epsilon_{n \bm k}},
\end{aligned}\end{equation}
and the $O(|\Delta|^2)$ term vanishes.
Similarly, the variation in the electron density with respect to the gap function is
\begin{equation} \begin{aligned}
\frac{\delta n( \bm r)}{\delta \Delta^*_{n \bm k}} 
&= \frac{\Delta_{n \bm k}}{\epsilon_{n \bm k}} 
\left( \frac{1-2n(\epsilon_{n \bm k})}{2\epsilon_{n \bm k}} + \frac{\partial n(\epsilon_{n \bm k})}{\partial \epsilon_{n \bm k}} \right) |\phi_{n \bm k}|^2   \\
& \quad +  O(|\Delta|^3),
\label{eq:density_variation}
\end{aligned}\end{equation}
which is negligible by the same argument.
Strictly the chemical potential, which is fixed by the electron density, Eq.~\ref{eq:renormalised_density}, also depends upon $\Delta_{n \bm k}$.
However the results above also imply that this dependence is negligible, since 
\begin{equation}
\frac{\delta \mu}{\delta \Delta^*_{n \bm k}} 
= \int d \bm r \ \frac{\partial \mu}{\partial n(\bm r)}  \frac{\partial n(\bm r)}{\partial \Delta^*_{n \bm k}}.
\end{equation}
		
The stationary equations  for $\phi_{n \bm k}( \bm r)$ are modified Kohn-Sham equations,
\begin{equation} 
\left(  -\frac{\nabla^2}{2} + v_{_{\text{KS}}}( \bm r) \right) \phi_{n \bm k} ( \bm  r)  
+  \frac{\delta F^{\Delta}}{\delta \phi^*_{n \bm k}( \bm r )} = \epsilon_{n \bm k}\phi_{n \bm k} ( \bm r).
\end{equation}
The term $\delta F^{\Delta}/\delta \phi^*_{n \bm k}( \bm r) $ can be identified as an  $O(|\Delta|^2)$ correction to the Kohn-Sham potential.
We make a key approximation which is to neglect this correction, which leaves the normal state Kohn-Sham equations, 
Eq.~\ref{eq:KS_equations}.
These are now constrained by the renormalised electron density, Eq. \ref{eq:renormalised_density}, and therefore the Kohn-Sham spectrum depends implicitly on the gap function.
This is similar to the adiabatic approximation of time-dependent DFT: the superconductivity is assumed to be a sufficiently weak perturbation on the non-superconducting system such that it always remains in its equilibrium state, only the equilibrium density is modified.
It should be noted that the same approximation is made in SCDFT \cite{Lueders2005}.
				
The only contribution to the stationary equations for $\Delta_{n \bm k}$ comes from the minimum of $F^{\Delta}$.
We find the superconducting gap equation,		
\begin{equation}\begin{aligned}
&0 = \Delta_{\alpha  \bm  k}\left(1 - \frac{\partial \Sigma_{\alpha  \bm  k}(\epsilon_{\alpha  \bm  k})}{\partial \epsilon_{\alpha  \bm  k}} 
\Bigg|_{\epsilon_{\alpha  \bm  k}=0} \right) \\
&+ \sum_{ \bm  k^\prime} \Delta_{\alpha  \bm  k^\prime} 
\frac{1- 2 n(\epsilon_{\alpha  \bm  k^\prime})}{2\epsilon_{\alpha  \bm  k^\prime}} 
\Big( \Lambda^C_{P \ \alpha  \bm  k, \alpha^\prime  \bm  k^\prime}(0) 
+ \Lambda^S_{P \ \alpha  \bm  k, \alpha^\prime  \bm  k^\prime}( 0) \Big),
\label{eq:QOBD_DFT_gap_eq}
\end{aligned}\end{equation}
where the effective interactions, $\Lambda^{C,S}$, and one-electron self-energy, $\Sigma$, are defined as in Eq.~\ref{eq:QOBD_SE_SF_propagators}.
By virtue of working with the $O(|\Delta|^2)$ free energy, this is the linearised gap equation that can only be used to determine the critical temperature.
To allow a finite gap function, the free energy must be expanded to higher order in $\Delta_{\bm k}$.
It was noted in \cite{Lueders2005} that the full non-linear gap equation contains many divergences and pathologies, and that a more stable way to allow a finite gap function -- that also yields accurate results -- is to allow the superconductivity to feed back onto the pairing susceptibility. 
This is achieved by making the replacement,
\begin{equation}
\frac{1 - 2n(\epsilon_{n \bm k^\prime}) }{2\epsilon_{n \bm k^\prime}} \rightarrow \frac{1 - 2n(\xi_{n \bm k^\prime}) }{2\xi_{n \bm k^\prime}}, 
\label{eq:SC_replacement}
\end{equation}
in the pairing term, to give the ‘partially linearised gap equation'.
This approach is also taken here.
One final simplification is made, which is to take an ansatz for the gap function that describes pairing in a particular angular 	momentum channel.
This means setting
\begin{equation}
\Delta_{\alpha \bm k} = |\Delta| d_{\bm k},
\label{eq:SC_ansatz}
\end{equation}
where $d_{\bm k}$ is a form factor.
This greatly simplifies the calculation by replacing an entire functional optimisation with the optimisation of a single 			number, $|\Delta|$.

\begin{figure}
\includegraphics[scale=0.75]{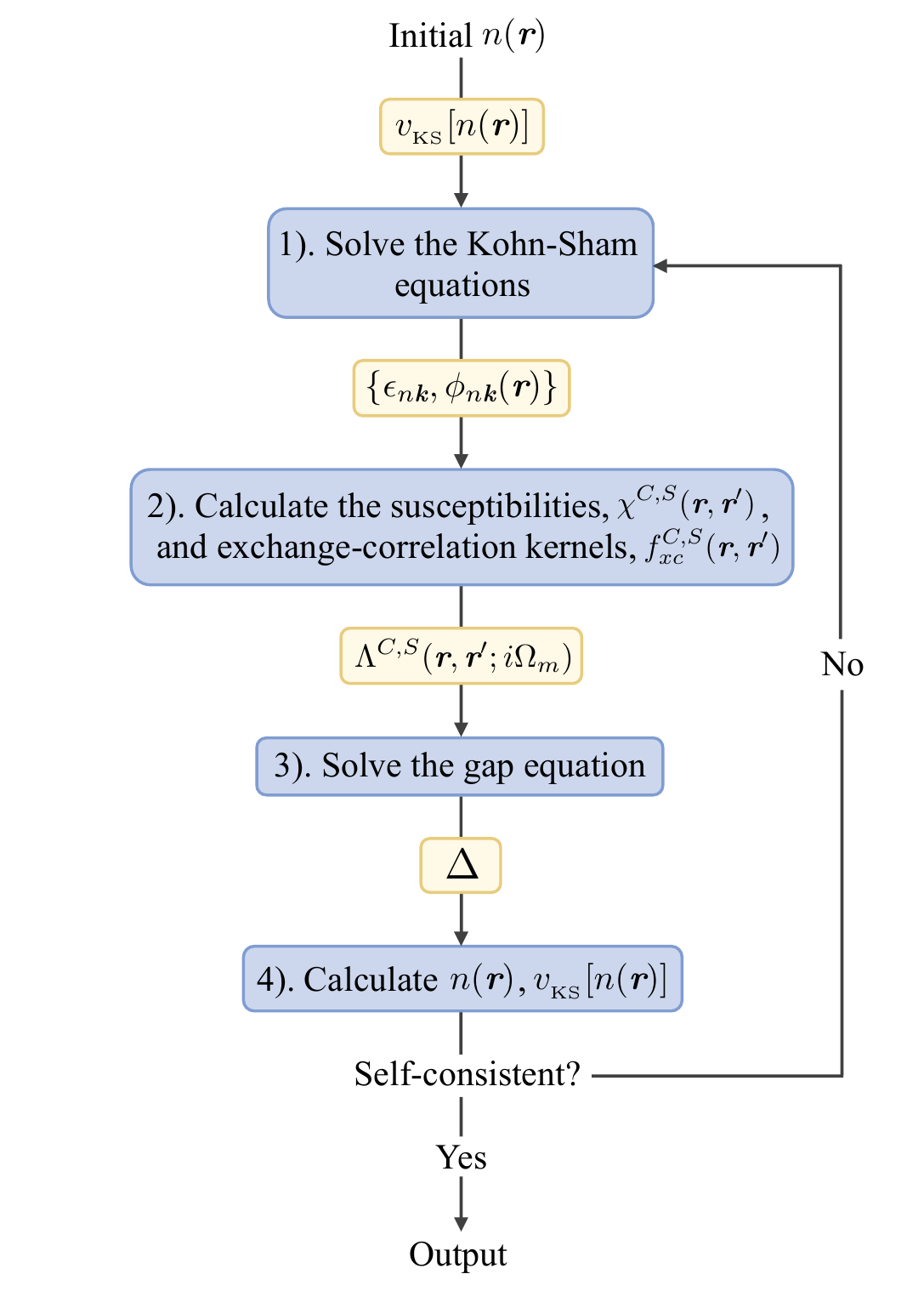}
\caption{\label{fig:SC_implementation} \textit{An implementation scheme for the superconducting QOBD-DFT framework.} 
The Kohn-Sham equations and the gap equation of the superconducting Kohn-Sham system are solved self-consistently with one another.
These equations are coupled implicitly by the electron density, which is modified by the superconducting order.
The self-consistent scheme is structurally identical when considering spin nematic order, only a self-consistency equation for the spin nematic order parameter, Eq.~\ref{eq:nematic_gap}, is solved in place of the superconducting gap equation, and the electron density is modified by the spin nematic order as in Eq.~\ref{eq:nematic_density}.}
\end{figure}
	
The superconducting Kohn-Sham system is therefore solved by solving the normal state Kohn-Sham equations and the 		superconducting gap equation self-consistently with one another.
Of course, this is not an easy computational task as both of these sets of equations must themselves be solved self-consistently.
An implementation scheme is presented in Fig. \ref{fig:SC_implementation}, for the case of a paramagnetic and non-spin-orbit-coupled system. 
Within this calculation, the DFT susceptibilities are calculated using their Dyson equations (as opposed to the 				Sternheimer equations) using the output of a spin-degenerate DFT calculation.
The steps are as follows:

\begin{enumerate}
\item The Kohn-Sham potential is constructed for an input electron density, and the Kohn-Sham equations,
Eq.~\ref{eq:KS_equations}, are solved for the spectrum, $\{\epsilon_{n\bm k}, \phi_{n\bm k} \}$. 
\item Using these eigenstates, the exchange-correlation kernels, $f_{xc}^{C,S}$, and DFT charge and spin susceptibilities, $\chi^{C,S}$,  are calculated as in Eq.~\ref{eq:DFT_kernels} and Eq.~\ref{eq:sus_Dyson} respectively, and the charge and spin fluctuation propagators are constructed, Eq.~\ref{eq:DFT_fluct_propagators}.
The electronic self-energy, Eq.  \ref{eq:QOBD_SE_SF_propagators}, is also calculated and its energy gradient is evaluated at the chemical potential.
\item These results are entered into the partially linearised gap equation (Eq.~\ref{eq:QOBD_DFT_gap_eq} with Eq.~\ref{eq:SC_ansatz} and the replacement described in Eq.~\ref{eq:SC_replacement}) which is solved self-consistently for $|\Delta|$.
\item A new, renormalised electron density is calculated using Eq.~\ref{eq:renormalised_density}. This used to identify a new input density if a convergence condition is not satisfied.
\end{enumerate}
			
\subsection{\label{subsection:spin_nematic}An implementation scheme for spin nematic order}

The QOBD-DFT approach may also be used to introduce arbitrary fluctuation-driven spin order into spin-degenerate DFT.
Here, spin nematic order is considered as described previously, i.e. corresponding to a finite value of $\langle \hat {\bm R}^t_{l=2,m, \bm q=0} \rangle$, for a particular $m$.
The order is introduced into a single band that exists at the chemical potential, labelled $\alpha$, by introducing the variational terms, 
\begin{equation}
S_{var}[\eta] = \eta \sum_{k \sigma} \ \sigma d_{\bm k} c^\dagger_{\alpha k\sigma} c_{\alpha k\sigma},
\label{eq:nematic_var}
\end{equation}
into the action of the interacting Kohn-Sham system.
The electronic dispersion becomes modified by the order as 
$\xi_{n \bm k \sigma}(\eta) = \epsilon_{n \bm k} - \sigma d_{\bm k} \eta \delta_{n \alpha} $, where $\epsilon_{n \bm k}$ has absorbed   the chemical potential.

In the presence of spin nematic order the spin densities are no longer equal,  
\begin{equation} \begin{aligned}
n_\sigma(\bm r) 
&=  \sum_{n\bm k} n(\xi_{n\bm k\sigma }) |\phi_{n\bm k\sigma}(\bm r)|^2  \\
&= n(\bm r) + \Delta n_\sigma [\eta](\bm r) \\ 
&= n(\bm r)- \sigma \Delta m[\eta] (\bm r) +  \Delta n[\eta] (\bm r). 
\label{eq:nematic_spin_densities}
\end{aligned}\end{equation}
The changes in the magnetisation and electron densities due to the order are indicated in the final line.
The introduction of the spin order into the Kohn-Sham system effectively sets a particular ansatz for the spin densities.
This ansatz, defined by the electron density, $n(\bm r)$, and the spin nematic order parameter, $\eta$, 
yields a space of spin densities, over which the free energy must be minimised, that is much reduced from the full space of 
spin density matrices of spin-DFT.
In spite of this, the full stationary equations for $n(\bm r)$ and $\eta$ remain highly coupled.
Of course, spin nematic order along the $z$ axis that is considered here is a particularly simple case; spin nematic order in other directions/channels will lead to more complicated modifications to the full spin density matrix, $n_{\sigma \sigma^\prime}(\bm r)$ \cite{Hannappel2016}.

Applying the QOBD procedure gives the $O(\eta^2)$ free energy,
\begin{equation} \begin{aligned}
&F[\eta] = F[\eta= 0] \\ 
&- \eta^2 \sum_{\bm k} d_{\bm k}^2 \ n^\prime (\epsilon_{\alpha \bm k})
\left(1 - \frac{\partial \Sigma_{\alpha \bm k} (\epsilon_{\alpha \bm k}) }{\partial \epsilon_{\alpha \bm k}} 
\Bigg|_{\epsilon_{\alpha \bm k} = 0} \right) \\
&- \eta^2 \sum_{\bm k \bm k^\prime} d_{\bm k} d_{\bm k^\prime} n^\prime (\epsilon_{\alpha \bm k}) n^\prime (\epsilon_{{\alpha \bm k}^\prime}) \Big(\Lambda^C_{ \alpha \bm k, \alpha \bm k^\prime}(0) + \Lambda^S_{ \alpha \bm k, \alpha \bm k^\prime}(0) \Big).
\label{eq:expanded}
\end{aligned}\end{equation}
This is the free energy functional of a spin nematic system, whose $\eta$-dependent terms describe corrections to the exchange-correlation functional due to the order. As in the superconducting case, this is an orbital-dependent functional.
The result depends on the self-energy,
\begin{equation} \begin{aligned}
&\Sigma_{n \bm k}(i\omega_n) =   \sum_{n^\prime \bm k^\prime } \sum_{i \omega_{n^\prime}} 
G^0_{n^\prime \bm k^\prime}(i\omega_{n^\prime})
\ \Lambda^S_{ n \bm k, n^\prime \bm k^\prime} (i\omega_n - i\omega_{n^\prime}), 
\end{aligned}\end{equation}
and the spin fluctuation propagators of the non-ordered state,
\begin{multline}
\Lambda^{S,C}_{ n \bm k, n^\prime \bm k^\prime} (i\Omega_m) = \int d\bm r d \bm r^\prime \\
\phi^*_{n^\prime \bm k^\prime}(\bm r)  \phi^*_{n\bm k}(\bm r^\prime) 
\ \Lambda^{S,C}(\bm r, \bm r^\prime; i\Omega_m)\ \phi_{n^\prime \bm k^\prime}(\bm r^\prime) \phi_{n \bm k}(\bm r) ,
\end{multline}
where $\Lambda^{S,C}(\bm r, \bm r^\prime; i\Omega_m)$ are given by Eq. \ref{eq:DFT_fluct_propagators}.
Since the nematic order parameter does not enter the susceptibilities, the fluctuation propagators may be calculated within spin-degenerate DFT.	
As in the superconducting case, $F[\eta= 0]$ is identified with the DFT thermodynamic potential, $\Omega^{DFT}$ (Eq. \ref{eq:DFT_free_energy}).
The $O(\eta^2)$ free energy of Eq.~\ref{eq:expanded} can only be used to calculate the ordering temperature. In such a “one-shot" calculation, the $O(\eta^2)$ coefficient in the free energy is evaluated using the spectrum of the spin-degenerate Kohn-Sham equations, Eq. \ref{eq:KS_equations}.	

A finite order parameter requires expanding the free energy to higher orders in $\eta$. 
This might be achieved, in manner similar to the superconducting case, by allowing the nematic order to modify the density of states at the Fermi level, i.e. \textit{via} the replacements, $n^\prime (\epsilon_{\alpha \bm k}) \rightarrow n^\prime (\xi_{\alpha \bm k \sigma})$.
The stationary equations can then be determined in a small-$\eta$ approximation scheme equivalent to that applied to the superconducting system.
The spin order modifies the electron density as 
\begin{equation}
n(\bm r) =  \frac{1}{2} \sum_{n\bm k \sigma} n(\xi_{n\bm k\sigma }) |\phi_{n\bm k\sigma}(\bm r)|^2.
\label{eq:nematic_density}
\end{equation}
Assuming that $\delta n(\bm r)/ \delta \phi^*_{n\bm k}(\bm r)$ depends weakly on $\eta$, and neglecting the $O(\eta^2)$ correction to $v_{_{\text{KS}}}(\bm r)$, then the spectrum of the spin nematic Kohn-Sham system is the solution of the spin-degenerate Kohn-Sham equations, Eq. \ref{eq:KS_equations}.
Similarly, minimising the free energy with respect to $\eta$ and neglecting the small, implicit dependence \textit{via} $n(\bm r)$ yields a self-consistency equation for $\eta$ akin to the superconducting gap equation,
\begin{equation} \begin{aligned}
0&= \sum_{\bm k } \sum_{\sigma} d_{\bm k}^2 \ \frac{n^\prime (\xi_{ \alpha \bm k \sigma})}{2}
\left(1 - \frac{\partial \Sigma_{ \alpha \bm k} (\epsilon_{ \alpha \bm k}) }{\partial \epsilon_{ \alpha \bm k}} \Bigg|_{\epsilon_{\alpha  \bm k} = 0} \right) \\
&-  \sum_{\bm k \bm k^\prime} \sum_{\sigma \sigma^\prime} \ d_{\bm k} d_{\bm k^\prime} \ \frac{n^\prime (\xi_{\alpha \bm k \sigma}) n^\prime ( \xi_{\alpha \bm k^\prime \sigma^\prime}) }{2} \\
&\quad\quad\quad\quad \Big( \Lambda^C_{\alpha \bm k, \alpha  \bm k^\prime}(0) \delta_{\sigma \sigma^\prime}  + \Lambda^S_{ \alpha \bm k, \alpha  \bm k^\prime}(0) \delta_{\bar{\sigma} \sigma^\prime} \Big), 
\label{eq:nematic_gap}
\end{aligned} \end{equation}
where $\bar\sigma$ indicates a flipped spin.
The self-consistent solving scheme is then structurally identical to that of the superconducting system, presented in 				Fig.~\ref{fig:SC_implementation}. 
Eq.~\ref{eq:nematic_gap} is solved self-consistently with the normal state Kohn-Sham equations, Eq.~\ref{eq:KS_equations}, which are coupled by the renormalised electron density, Eq.~\ref{eq:nematic_density}.
In this scheme, the spin nematic order explicitly modifies the (spin-resolved) density of states, but not the fluctuation properties of the system.
The fluctuation propagators that enter Eq.~\ref{eq:nematic_gap} are those of the spin-degenerate (non-ordered) system, only now they are calculated using the Kohn-Sham spectrum and electron density that are renormalised by the spin nematic order.

\section{Proof of principle}

A proof of principle of the approach is best achieved by studying the solution to a simple toy model, which we take to be the single band Hubbard model.
This model has already been considered within the QOBD framework in Sections \ref{subsection:QOBD_SC} and \ref{subsection:QOBD_nematic}.
For clarity we give an explicit re-statement of this treatment within the framework of density functional theory. 
 
The density functional of the model at the level of the treatments in \cite{Karahasanovic2012, Conduit2013, Hannappel2016} is given by the derivative of the Abrikosov-Khalatnikov free energy \cite{Abrikosov1958} presented here in Eq.~\ref{eq:QOBD_ferro}
with respect to the charge density. This corresponds to a self-consistent 
re-summation of second order corrections calculated around a uniform charge density. 
The sequence of works in Refs.~\cite{Karahasanovic2012, Conduit2013, Hannappel2016} describe spatial-modulated magnetic, superconducting and spin-nematic order in this background. The corresponding corrections to the free energy due to superconducting and spin-nematic order are given by
\begin{equation} \begin{aligned}
F[\Delta] &=  \sum_{\bm k}  |\Delta_{\bm k}|^2 \ \frac{1-2n (\epsilon_{\bm k})} {2\epsilon_{\bm k}}  
\left(1 - \frac{\partial \Sigma_{\bm k}(\epsilon_{ \bm k})}{\partial \epsilon_{\bm k}} 
\Bigg|_{\epsilon_{ \bm k}=0} \right) \\ 
&+  \sum_{\bm k \bm q} \bar\Delta_{\bm k} \Delta_{\bm k + \bm q}  \frac{1- 2n(\epsilon_{\bm k})}{2\epsilon_{\bm k}} \frac{1- 2 n(\epsilon_{\bm k + \bm q})} {2\epsilon_{\bm k + \bm q}}  \Lambda_{\bm q}(0), \\
F[\eta] =  &- \eta^2 \sum_{\bm k} d_{\bm k}^2 \ n^\prime (\epsilon_{ \bm k})
\left(1 - \frac{\partial \Sigma_{\bm k} (\epsilon_{ \bm k}) }{\partial \epsilon_{\bm k}} 
\Bigg|_{\epsilon_{ \bm k} = 0} \right) \\
&- \eta^2 \sum_{\bm k \bm q} d_{\bm k} d_{\bm k + \bm q} n^\prime (\epsilon_{ \bm k}) n^\prime (\epsilon_{\bm k + \bm q}) \Lambda_{\bm q}(0),
\label{eq:expanded}
\end{aligned}\end{equation}
where $\Lambda_{\bm q}(0) = U^2 \chi_{\bm q}(0)$ and $\epsilon_{\bm k} = -t \left(\cos k_x + \cos k_y \right)$, in terms of the Hubbard and hopping parameters of the model, $U$ and $t$.
Corrections due to magnetic order have non-analytic dependences as a function of the magnitude of  the magnetic order \cite{Belitz1997, Belitz1999} and its wave-vector \cite{Conduit2009} whose resummation was carried out in Ref.~\cite{Pedder2013}.
Interactions between these order parameters is  calculated through higher order terms in the expansion of 
$-T \sum_{\bm k, \sigma} \ln (1+ \exp[-\xi_{\bm k,\sigma}(N,M,\Delta, \eta)/T] )$.

The optimisation of these free energy corrections at small filling recovers -- by construction -- the phase diagrams presented in previous works \cite{Conduit2013,Karahasanovic2012,Hannappel2016}, which we refer the reader back to.
These studies provide a proof of principle of the theory.
An explicit inclusion of the QOBD free energy corrections in a density functional code will have the benefits of bringing \textit{ab initio} numerical accuracy to the results without an undue computational cost of a full spin- or superconducting density functional optimisation.
This work lays out the formal basis for such an inclusion.
		
\section{Summary and outlook}

In summary, QOBD-DFT is a framework for treating arbitrary electronic order that is stabilised by collective spin and 	charge fluctuations within density functional theory, from the platform of a local or semi-local approximation to exchange and correlation.
This is achieved by applying the quantum order-by-disorder approach to dynamically stabilised order to the Kohn-Sham system of spin-degenerate DFT.
By working with a variational order parameter, order may be investigated in a particular channel in a controlled manner, and by means of a functional optimisation over a reduced space of densities compared to full spin- or superconducting DFT.
This approach is made fully first-principles by calculating the propagators of spin and charge fluctuations within DFT.
Schemes have been presented for the examples of fluctuation-driven superconductivity, whose result is equivalent to an existing, formally-derived superconducting DFT, and spin nematic order.
		
An attractive strength of this approach lies in its ability to consider an arbitrary number of orders simultaneously and on equal footing.
An exciting prospect is to apply this method to study correlated metals in which multiple collective orders are in close competition, as  exemplified by the interplay of magnetic, nematic and superconducting orders in the iron pncitides and iron chalcogenides \cite{Fernandes2022}.
Then, working with an order parameter is a great advantage.
Multiple orders can be considered together in a very controlled manner by choosing ansatz based on symmetry considerations, and the resulting optimisation problem is as simple as possible.
This is reflected in the success that the QOBD approach has had in describing the ordered phases that appear in the vicinity of itinerant quantum critical points \cite{Karahasanovic2012, Conduit2013, Hannappel2016}.
Combining this approach with DFT introduces first-principles material-specific information, for minimal computational expense, whilst enjoying the benefits of an order parameter theory.

Although the implementation of this framework is beyond the scope of this publication, it is expected that this scheme can be achieved as a straightforward extension of existing DFT codes.
This could be achieved using CasPyTep \cite{Corbett2015}, for example, which is a Python interface to the CASTEP plane wave DFT code \cite{Clark2005}.
This interface allows all of the quantities required for a QOBD-DFT calculation to be extracted from and re-entered into a DFT calculation in a particularly straightforward manner.
In this way, DFT codes may be adapted to selectively consider fluctuation-driven electronic orders.

		
\begin{acknowledgments}
The authors benefited from stimulating discussions with James Kermode, Frank Krüger, Elliot Christou, Andrew James and David Bowler.
This work has been supported by EPSRC through grant EP/P013449/1.
\end{acknowledgments}

\appendix
\section{\label{Appendix:SC}QOBD for the spin-singlet superconductor}

Fluctuation-driven spin-singlet superconductivity is introduced into the Hubbard model, within the variational perturbation theory implementation of the quantum order-by-disorder framework laid out in Section \ref{section:QOBD}, using the variational terms, 
\begin{equation} 
S_{var}(\Delta) =  -\sum_k \Big( \Delta_{\bm k} c^\dagger_{k\uparrow} c^\dagger_{-k\downarrow} 
+ \bar{\Delta}_{\bm k} c_{-k\downarrow}c_{k\uparrow} \Big).
\end{equation}
A 4-momentum notation is applied, $k = (i\omega_n, \bm k)$ etc.
Adding and subtracting these terms to the action of the Hubbard model, Eq. \ref{eq:Hubbard_action}, gives 
$S = S_{0}^\prime + S_{int}^\prime$, where 	
\begin{equation} \begin{aligned}
S_{0}^\prime &= \sum_{k\sigma} c^\dagger_{k\sigma} (-i\omega_n + \epsilon_{\bm k})  c_{k\sigma} \\
&  \ + \sum_k \Big( \Delta_{\bm k} c^\dagger_{k\uparrow} c^\dagger_{-k\downarrow} 
+ \bar{\Delta}_{\bm k} c_{-k\downarrow}c_{k\uparrow} \Big), \\  
S_{int}^\prime &= - \sum_k \Big( \Delta_{\bm k} c^\dagger_{k\uparrow} c^\dagger_{-k\downarrow} 
+ \bar{\Delta}_{\bm k} c_{-k\downarrow}c_{k\uparrow} \Big) \\
& \  + U \sum^\prime_{kk^\prime pp^\prime} c^\dagger_{k \uparrow} c^\dagger_{p \downarrow} c_{p^\prime \downarrow}			c_{k^\prime \uparrow}.
\end{aligned} \end{equation}
The prime over the sum indicates the presence of a 4-momentum-conserving delta-function, $\delta(k + p - k^\prime - p^\prime)$, and the chemical potential is absorbed into $\epsilon_{\bm k}$ for clarity.
The propagator of the superconducting non-interacting system, $\mathcal G(\Delta)$, is identified from its partition function,
\begin{equation}
Z_0[\Delta]= \int D[\bar{c}, c]  e^{-S_0^\prime} =  e^{\text{Tr} \ln \mathcal G^{-1}(\Delta)}.
\end{equation}
In the basis of Nambu-Gorkov spinors, $\underline \psi_k^\dagger = (c^\dagger_{k\uparrow} \ c_{-k \downarrow} )$, we find 
$\mathcal G^{-1}_k (\Delta_{\bm k} ) = -i \omega_n \sigma_0 + \underline{\underline{h}}_{\bm k}$, 
where 
$ \underline{\underline{h}}_{\bm k} = \epsilon_{\bm k} \sigma_z +  \left( \Delta_{\bm k} + \bar \Delta_{\bm k} \right) \sigma_x /2 + i  \left( \Delta_{\bm k} - \bar \Delta_{\bm k} \right) \sigma_y /2$, in terms of Pauli matrices, $\sigma_i$.
Inverting this result gives the Nambu-Gorkov propagator (Eq.~\ref{eq:NG_GF}) whose elements are the normal and anomalous electronic propagators,
\begin{equation} \begin{aligned}
G_{ \bm k \sigma}(i\omega_n) &= -\langle c_{{\bm k} \sigma}(i\omega_n) c^\dagger_{{\bm k} \sigma} (i\omega_n)\rangle  \\
&= \frac{|u_{\bm k}|^2}{i\omega_n - \xi_{\bm k}} + \frac{|v_{\bm k}|^2}{i\omega_n + \xi_{\bm k}}, \\ 
F_{\bm k}(i\omega_n) &= - \langle  c_{ {\bm k} \uparrow} (i\omega_n)c_{ -{\bm k} \downarrow} (i\omega_n)\rangle \\
&=  \frac{\Delta_{\bm k}}{(i\omega_n - \xi_{\bm k} )(i\omega_n + \xi_{\bm k})}.
\label{eq:full_propagators}
\end{aligned}\end{equation}
These have been written in terms of the electron and hole amplitudes, 
$u_{\bm k} = \sqrt{\left(1 + \epsilon_{\bm k}/\xi_{\bm k} \right)/2}$ and 
$v_{\bm k} = e^{i\delta_{\bm k}} \sqrt{ \left(1 - \epsilon_{\bm k}/\xi_{\bm k} \right)/2}$, and the Bogoliubov dispersion, $\xi_{\bm k} = \sqrt{\epsilon_{\bm k}^2 + |\Delta_{\bm k}|^2}$.
These amplitudes define the Bogoliubov transformation, $\alpha_{{\bm k} \sigma}^\dagger = u_{\bm k} c^\dagger_{{\bm k} \sigma} + \sigma \bar{v}_{\bm k} c_{-{\bm k}-\sigma}$, that diagonalises the non-interacting Hamiltonian of the superconducting system.
The free energy is
\begin{equation} \begin{aligned}
F &= F_0 \ + \ F_1 \ + \ F_2 \\ 
&=  - T \ \text{Tr} \ln \mathcal G^{-1}(\Delta) \ + \  T \langle S^\prime_{int} \rangle_{0,c} \ - \  \frac{T}{2} \langle S_{int}^{2} \rangle_{0,c},
\end{aligned} \end{equation}
where terms are labelled according to their order in $U$ for ease of analysis. 
This result is retained up to quadratic order in the superconducting order parameter. 

We now evaluate the free energy term-by-term.
$F_0$ may be evaluated directly using $\mathcal G^{-1}_k $. After evaluating the Matsubara frequency sum, we find 
\begin{equation} \begin{aligned}
F_{0} &=  - 2T \sum_{\bm k} \ \ln \left(2 \cosh (\xi_{\bm k}/(2T))\right) \\ 
&= F_{0}[\Delta=0] - \sum_{\bm k} \ |\Delta_{\bm k}|^2 \ \frac{1- 2 n_{\bm k}}{2\epsilon_{\bm k}} + O(|\Delta|^4),
\label{append_eq:F0}
\end{aligned} \end{equation}
where the result is expanded in $\Delta$ in the second line, and $n_{\bm k} \equiv n(\epsilon_{\bm k})$.
The remaining terms, $F_1$ and $F_2$, are evaluated by first evaluating the expectation values \textit{via} Wick contractions of the electronic fields, which are expressed in terms of the electronic correlators, Eq.~\ref{eq:full_propagators}, before carrying out the Matsubara frequency sums to leave a result expressed as momentum integrals. 
Instead of expanding the result to $O(\Delta^2)$ at the end of the calculation, it is most efficient to expand the electronic propagators in $|\Delta|^2$ before performing the Matsubara frequency summations. 
Doing so gives 
\begin{equation} \begin{aligned}
G_{\bm k\sigma}(i\omega_n) 
&= G^0_{\bm k}(i\omega_n)  +  \bar{\Delta}_{\bm k} G^0_{\bm k}(i\omega_n) F^0_{\bm k}(i\omega_n) + O(|\Delta|^4),  \\ 
F_{\bm k}(i\omega_n) &= F^0_{\bm k}(i\omega_n) +  O(|\Delta|^3),
\label{eq:leading_order_propagators}
\end{aligned}\end{equation}
in terms of the non-interacting propagator of the non-superconducting system, $G^0_{{\bm k} \sigma}(i\omega_n) = (i\omega_n - \epsilon_{\bm k} )^{-1}$, and the leading order anomalous propagator,
$F^0_{\bm k}(i\omega_n) \equiv -\Delta_{\bm k} \ G^0_{\bm k}(i\omega_n) G^0_{\bm k}(-i\omega_n)$. 
We find the $O(U)$ free energy,
\begin{equation} \begin{aligned}
F_1  &= F_1[\Delta = 0] \ + \ 2 \sum_{\bm k} \ |\Delta_{\bm k}|^2 \ \frac{1-2n_{\bm k}}{2\epsilon_{\bm k}}  \\
&+ U \sum_{{\bm k}{\bm k}^\prime} \ \bar{\Delta}_{\bm k} \Delta_{{\bm k}^\prime} 
\frac{1-2n_{\bm k}}{2\epsilon_{\bm k}} \frac{1-2n_{{\bm k}^\prime}}{2\epsilon_{{\bm k}^\prime}}.
\label{append_eq:F1}
\end{aligned} \end{equation}
The free energy of the static ordered state may be assembled as $F_{_{SO}} =  F_0 + F_1$.

\begin{figure}
\includegraphics[scale=1.1]{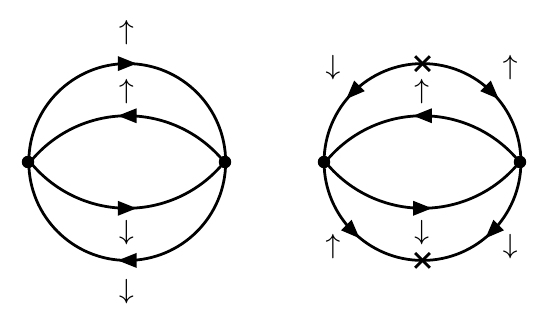}
\caption{\label{fig:Feyn_diags} \textit{Feynman diagram representations of $F_{renorm}$ (left) and $F_{pair}$ (right) in Eq. 		\ref{eq:feyn_diags_Fs}, with the spin structure indicated.} On the left, all lines denote electronic propagators, $G$; on the 			right, the top and bottom lines denote $F^\dagger$ and $F$, all of which are defined in Eq. \ref{eq:full_propagators}. }
\end{figure}

The free energy of fluctuations is given by $F_2$. Evaluating all Wick contractions gives 
\begin{equation} \begin{aligned}
F_2
&= F_{renorm} \ + \ F_{pair} \\
&=  -\frac{U^2}{2} \sum^\prime_{kk^\prime pp^\prime} 
 \left(  G_{k\uparrow}  G_{p \downarrow } G_{p^\prime \downarrow} G_{k^\prime \uparrow}
 \ + \ 2 \ F^\dagger_k  G_{p \uparrow} G_{ p^\prime \downarrow } F_{k^\prime } \right).
\label{eq:feyn_diags_Fs}
\end{aligned} \end{equation}
This is partitioned into two terms: the free energy due to field renormalisation and the free energy of pairing, which have useful Feynman diagrams, shown in Fig. \ref{fig:Feyn_diags}.
Expanding the propagators as in Eq. \ref{eq:leading_order_propagators} and keeping terms to $O(|\Delta|^2)$ gives
\begin{equation} \begin{aligned}
&F_{renorm} =  U^2 \sum_{\sigma}  \sum_{\bm k \bm k^\prime} \sum_{i\omega_n i\omega_n^\prime}  
G^0_{\bm k^\prime \bar \sigma}(i\omega_{n^\prime}) \chi^0_{\bm k \bm k^\prime}(i\omega_n - i\omega_{n^\prime}) \\
&\quad \quad \quad\quad \quad \quad \left[ G^0_{\bm k \sigma}(i\omega_n) + \bar{\Delta}_{\bm k} G^0_{\bm k \sigma}(i\omega_n) F^0_{\bm k \sigma}(i\omega_n) \right],  \\ \\ 
&F_{pair} =  U^2 \sum_{\bm k\bm k^\prime} \sum_{i\omega_n i\omega_n^\prime}
F^{0 \dagger}_{\bm k}(i\omega_n) \chi^0_{\bm k \bm k^\prime}(i\omega_n - i \omega_{n^\prime} )  
F^0_{\bm k^\prime} (\omega_{n^\prime}).
\label{append_eq:FR_FP}
\end{aligned} \end{equation}
We have identified the non-interacting susceptibility 
\begin{equation} \begin{aligned}
&\chi^0_{\bm k \bm k^\prime} (i\omega_n - i\omega_{n^\prime} ) = - \sum_{\bm p \bm p^\prime} \sum_{i\omega_m}  \delta(\bm k + \bm p - \bm k^\prime - \bm p^\prime) \\
&\quad\quad\quad\quad\quad\quad G_{\bm p \uparrow }(i\omega_m + i\omega_n - i\omega_{n^\prime}) G_{\bm p^\prime \downarrow}(i\omega_{m}) \\
&= \sum_{\bm p\bm p^\prime}
\ \frac{n_{\bm p} - n_{\bm p^\prime } }
 {i\omega_n - i\omega_{n^\prime}  + \epsilon_{\bm p^\prime} - \epsilon_{\bm p}}
\ \delta(\bm k + \bm p - \bm k^\prime - \bm p^\prime).
\end{aligned}\end{equation}
These results will be simplified by taking an on-shell approximation, which assumes that the electrons that form Cooper pairs occupy states at the Fermi level.		

The pairing term is evaluated by first carrying out the frequency sums, giving
\begin{equation} \begin{aligned}
&F_{pair} =  U^2 \sum_{\bm k\bm k^\prime \bm p\bm p^\prime}  \delta(\bm k + \bm p - \bm k^\prime - \bm p^\prime) \ 
\frac{\bar \Delta_{\bm k}}{2\epsilon_{\bm k}} \frac{\Delta_{\bm k^\prime}}{2\epsilon_{\bm k^\prime}} \\ 
& \Bigg( \frac{n_{\bm k} n_{\bm p} (1-n_{\bm p^\prime})(1-n_{\bm k^\prime}) 
\ - \  (1-n_{\bm k})(1- n_{\bm p}) n_{\bm p^\prime} n_{\bm k^\prime} }
{\epsilon_{\bm k} + \epsilon_{\bm p} - \epsilon_{\bm p^\prime} - \epsilon_{\bm k^\prime} }  \\ 
&  + \frac{n_{\bm k} (1-n_{\bm p}) n_{\bm p^\prime} n_{\bm k^\prime} 
\ - \  (1-n_{\bm k})n_{\bm p} (1-n_{\bm p^\prime})(1- n_{\bm k^\prime}) }
{-\epsilon_{\bm k} + \epsilon_{\bm p} - \epsilon_{\bm p^\prime} - \epsilon_{\bm k^\prime} }  \\ 
&  +  \frac{(1-n_{\bm k}) n_{\bm p} (1-n_{\bm p^\prime})n_{\bm k^\prime}
 \ - \  n_{\bm k}(1- n_{\bm p}) n_{\bm p^\prime} (1-n_{\bm k^\prime}) }
{-\epsilon_{\bm k} + \epsilon_{\bm p} - \epsilon_{\bm p^\prime} + \epsilon_{\bm k^\prime} } \\ 
&  +  \frac{(1-n_{\bm k}) (1-n_{\bm p}) n_{\bm p^\prime} (1-n_{\bm k^\prime}) 
\ - \  n_{\bm k} n_{\bm p} (1-n_{\bm p^\prime}) n_{\bm k^\prime} }
{\epsilon_{\bm k} + \epsilon_{\bm p} - \epsilon_{\bm p^\prime} + \epsilon_{\bm k^\prime} }
 \Bigg).
\label{eq:full_pairing_free_energy}
\end{aligned} \end{equation} 
Now going on-shell, i.e. taking $\epsilon_{\bm k,\bm k^\prime} \rightarrow 0$ in the denominators within the brackets, gives 
\begin{equation} \begin{aligned}
F_{pair} = U^2\sum_{\bm k\bm k^\prime} \ \bar \Delta_{\bm k} \Delta_{\bm k^\prime} \ \frac{1- 2 n_{\bm k}}{2\epsilon_{\bm k}} 
\frac{1- 2 n_{\bm k^\prime}} {2\epsilon_{\bm k^\prime}} \ \chi^0_{\bm k\bm k^\prime} (0).
\label{eq:pairing_result_append}
\end{aligned}\end{equation}
The on-shell approximation can equivalently be applied during the frequency sums in Eq.~\ref{append_eq:FR_FP}.
This is achieved by assuming that the poles of the Green's functions of the pairing electrons dominate the sum, and neglecting the poles of the dynamical susceptibility, i.e.
\begin{equation} \begin{aligned}
&F_{pair} \approx U^2 \sum_{\bm k \bm k^\prime }
 \frac{\bar \Delta_{\bm k}}{2\epsilon_{\bm k}} \frac{\Delta_{\bm k^\prime}}{2\epsilon_{\bm k^\prime}} \\
&\Big(n_{\bm k} \chi^0_{\bm k \bm k^\prime} (\epsilon_{\bm k} - \epsilon_{\bm k^\prime})  n_{\bm k^\prime} - n_{\bm k} \chi^0_{\bm k \bm k^\prime} (\epsilon_{\bm k} + \epsilon_{\bm k^\prime}) (1 - n_{\bm k^\prime}) \\ 
&\quad - (1-n_{\bm k}) \chi^0_{\bm k \bm k^\prime} (-\epsilon_{\bm k} - \epsilon_{\bm k^\prime})  n_{\bm k^\prime} \\ 
&\quad +(1-n_{\bm k})\chi^0_{\bm k \bm k^\prime} (-\epsilon_{\bm k} + \epsilon_{\bm k^\prime}) (1-n_{\bm k^\prime}) \Big),
\end{aligned} \end{equation}
where $F^0_{\bm k}(i\omega_n) = \Delta_{\bm k} (G^0_{\bm k}(i\omega_n) + G^0_{\bm k}(-i\omega_n))/(2\epsilon_{\bm k})$ has been applied.
Approximating the susceptibilities by their values at the chemical potential (i.e. taking $\epsilon_{\bm k, \bm k^\prime} \rightarrow 0$ in $\chi^0_{\bm k \bm k^\prime}$) retrieves the on-shell result, Eq.~\ref{eq:pairing_result_append}.

The first term in $F_{renorm}$ is independent of $\Delta$ and is neglected for clarity before being reintroduced later.
The remaining term is rewritten as	
\begin{equation} \begin{aligned}
F_{renorm} =  2 \sum_{\bm k} \sum_{i \omega_n} \
\bar{\Delta}_{\bm k} \ G^0_{\bm k}(i\omega_n)  F^0_{\bm k}(i\omega_n) \Sigma_{\bm k}(i\omega_n),  
\label{eq:F_renorm_pre_antisymm}
\end{aligned} \end{equation}
in terms of the (one-loop) electronic self-energy, 	
\begin{equation} \begin{aligned}
\Sigma_{\bm k}&(i\omega_n) =  U^2  \sum_{\bm k^\prime} \sum_{i\omega_{n^\prime}} \ 
G^0_{\bm k^\prime}(i\omega_{n^\prime}) \ \chi^0_{\bm k \bm k^\prime} (i\omega_n - i\omega_{n^\prime}) \\ 
&=  U^2 \sum_{\bm k^\prime \bm  p \bm p^\prime} 
\frac{ n_{\bm p} (1 - n_{\bm p^\prime})(1 - n_{\bm k^\prime}) + (1-n_{\bm p})n_{\bm p^\prime} n_{\bm k^\prime} }
{i\omega_n + \epsilon_{\bm p} - \epsilon_{\bm p^\prime} - \epsilon_{\bm k^\prime}} \\
&\quad\quad\quad\quad\quad\quad \delta(\bm k + \bm p - \bm k^\prime - \bm p^\prime).
\end{aligned}\end{equation}
Next, the part of the self-energy that is symmetric in frequency is neglected, i.e.
$\Sigma_{\bm k}(i\omega_n) \approx  \Sigma^A_{\bm k}(i\omega_n) 
= \left( \Sigma_{\bm k}(i\omega_n) - \  \Sigma_{\bm k}(-i\omega_n) \right)/2.$
We now find that part of the summand in Eq. \ref{eq:F_renorm_pre_antisymm} is antisymmetric in frequency and 				vanishes under the sum. We are left with 
\begin{equation} \begin{aligned}
F_{renorm} = 2 \sum_{\bm k} \sum_{i\omega_n}
\frac{ |\Delta_{\bm k}|^2}{2\epsilon_{\bm k}} \left( G^0_{\bm k}(i\omega_n) \right)^2  \Sigma^A_{\bm k}(i\omega_n).
\label{append_eq:F_renorm_sum}
\end{aligned} \end{equation}
The frequency sum is carried out, giving 
\begin{equation} \begin{aligned}
&F_{renorm} =  U^2 \sum_{\bm k \bm k^\prime \bm p \bm p^\prime} \delta(\bm k + \bm p - \bm k^\prime - \bm p^\prime) 
\frac{|\Delta_{\bm k}|^2}{2\epsilon_{\bm k}} \frac{\partial \ }{\partial \epsilon_{\bm k}} \\
&\Bigg( \frac { n_{\bm k} n_{\bm p} (1 - n_{\bm p^\prime}) (1 - n_{\bm k^\prime}) 
-  (1 - n_{\bm k})(1 - n_{\bm p} ) n_{\bm p^\prime} n_{\bm k^\prime}}					
{ \epsilon_{\bm k} + \epsilon_{\bm p} - \epsilon_{\bm p^\prime} - \epsilon_{\bm k^\prime} } \\ 
& - \frac
{  n_{\bm k}(1 - n_{\bm p} ) n_{\bm p^\prime} n_{\bm k^\prime} 
- (1 - n_{\bm k}) n_{\bm p} (1 - n_{\bm p^\prime}) (1 - n_{\bm k^\prime})  }								
{ -\epsilon_{\bm k} + \epsilon_{\bm p}  - \epsilon_{\bm p^\prime}  - \epsilon_{\bm k^\prime} } \Bigg).
\label{eq:full_renormalisation_free_energy}
\end{aligned}\end{equation}
Finally, the derivative is evaluated and an on-shell approximation is made, whereby the limit $\epsilon_{\bm k} \rightarrow 0$ 		is taken in the denominators, $\pm\epsilon_{\bm k} + \epsilon_{\bm p}  - \epsilon_{\bm p^\prime}  - \epsilon_{\bm k^\prime}$.
The result may then be written, 
\begin{equation} \begin{aligned}
F_{renorm}  
= -\sum_{\bm k}  |\Delta_{\bm k}|^2 \ \frac{1-2n_{\bm k}}{2\epsilon_{\bm k}}  \ 
\frac{\partial \Sigma_{\bm k}(\epsilon_{\bm k})}{\partial \epsilon_{\bm k}} \Bigg|_{\epsilon_{\bm k}=0}.
\label{eq:F_renorm_append}
\end{aligned}\end{equation}
As before, the on-shell approximation can equivalently be applied by assuming that the frequency sum in Eq.~\ref{append_eq:F_renorm_sum} is dominated by the pole of the Green's function, and neglecting the pole of the self-energy, i.e.
\begin{equation} \begin{aligned}
&\sum_{i\omega_n} \left( G^0_{\bm k}(i\omega_n) \right)^2  \Sigma^A_{\bm k}(i\omega_n) 
= \frac{\partial \ }{\partial \epsilon_{\bm k}} \bigg(\sum_{i\omega_n}  G^0_{\bm k}(i\omega_n) \Sigma^A_{\bm k}(i\omega_n) \bigg) \\
&=  \frac{1}{2} \frac{\partial \ }{\partial \epsilon_{\bm k}} \bigg( \sum_{i\omega_n} \left( G^0_{\bm k}(i\omega_n) - G^0_{\bm k}(- i\omega_n) \right)  \Sigma_{\bm k}(i\omega_n) \bigg) \\
&\approx \frac{1}{2} \frac{\partial \ }{\partial \epsilon_{\bm k}} \bigg( n_{\bm k} \Sigma_{\bm k}(\epsilon_{\bm k}) + (1- n_{\bm k}) \Sigma_{\bm k}(-\epsilon_{\bm k}) \bigg).
 \end{aligned} \end{equation}
Approximating the self-energy and its derivative by their values at the chemical potential retrieves the result in Eq.~\ref{eq:F_renorm_append}.
The full free energy is assembled as $F = F_{_{SO}} + F_{fl}$, where $F_{fl} = F_{renorm} + F_{pair}$, using Eqs. \ref{append_eq:F0}, \ref{append_eq:F1}, \ref{eq:pairing_result_append} and \ref{eq:F_renorm_append}, giving the result presented in Eq.~\ref{eq:free_energy_SC1}.
	
\section{\label{Appendix:nematic}QOBD for the spin nematic}

Spin nematic order is introduced into the Hubbard model by adding and subtracting the variational terms,
\begin{equation}
S_{var}(\eta) = \eta \sum_{k \sigma} \ \sigma d_{\bm k} c^\dagger_{k\sigma} c_{k\sigma},
\end{equation}
to the action, Eq.~\ref{eq:Hubbard_action}.
The propagator of the ordered non-interacting system can be identified straightforwardly as $G_{\bm k \sigma} (\eta) = (i\omega_n - \xi_{\bm k \sigma})^{-1}$, where $\xi_{\bm k \sigma} = \epsilon_{\bm k} - \sigma d_{\bm k} \eta$.
Evaluating the free energy as defined in Eq.~\ref{eq:QOBD_free_energy} gives
\begin{equation} \begin{aligned}
&F[\eta] = -T \sum_{ \bm k\sigma} \ln \left(1+ e^{-\xi_{\bm k\sigma} /T} \right) +  \sum_{\bm k\sigma} \sigma d_{\bm k} n_{\bm k\sigma} \\ 
&+ U \sum _{\bm k \bm p} n_{\bm k\uparrow} n_{\bm p \downarrow} + U^2 \sum_{\bm k\bm p\bm q} 
\frac{ n_{\bm k+\bm q \uparrow} n_{\bm p-\bm q \downarrow} (1- n_{\bm p \downarrow})  (1- n_{\bm k \uparrow})}
{\xi_{\bm k+\bm q \uparrow} + \xi_{\bm  p - \bm q \downarrow} - \xi_{\bm p \downarrow} - \xi_{\bm k \uparrow}} \\ 
&\quad \ \ = F_0 + F_{var} + F_1 + F_{fl},
\label{eq:append_nematic_F_1}
\end{aligned}\end{equation}	
where $n_{\bm k\sigma} \equiv n(\xi_{\bm k \sigma})$. Each term is labelled for ease of analysis.
A series of approximations are applied to this result that are equivalent to those made in the superconducting case:
$\eta$ is assumed to be small (such that we may expand to quadratic order); the self-energy is assumed to be anti-symmetric in frequency; and the electrons that order are assumed to exist at the Fermi level (permitting an on-shell approximation).

Expanding the free energy in Eq.~\ref{eq:append_nematic_F_1} to $O(\eta^2)$, term-by-term, gives
\begin{equation} \begin{aligned}
F_0&[\eta] = F_0[\eta=0] + \frac{\eta^2}{2} \sum_{\bm k\sigma}  d_{\bm k}^2 n^\prime(\epsilon_{\bm k}) + O(\eta^3), \\ 
F_{var} &[\eta] =-\eta^2  \sum_{\bm k\sigma} d_{\bm k}^2 n^\prime(\epsilon_{\bm k})+ O(\eta^3), \\ 
F_1&[\eta] = F_1[\eta=0] + \frac{U \eta^2}{2} \sum_{\bm k\bm k^\prime} 
\Big( d_{\bm k}^2 n^{\prime\prime}(\epsilon_{\bm k}) n(\epsilon_{\bm k^\prime}) \\
&+ d_{\bm k^\prime}^2 n(\epsilon_{\bm k}) n^{\prime\prime}(\epsilon_{\bm k^\prime}) 
-  2 d_{\bm k}  d_{\bm k^\prime} n^\prime (\epsilon_{\bm k})  n^\prime (\epsilon_{\bm k^\prime}) \Big) + O(\eta^3).
\end{aligned} \end{equation} 
The terms in $F_1$ that are proportional to $d_{\bm k}^2$ vanish when taken on-shell (i.e. when $\epsilon_{\bm k,\bm k^\prime}\rightarrow 0$) since 
$n^{\prime\prime}(\epsilon_{\bm k})$ is an antisymmetric function which vanishes at $\epsilon_{\bm k}=0$.
Evaluating $F_{_{SO}} = F_0 + F_{var} + F_1$ gives 
\begin{equation} \begin{aligned}
F_{_{SO}}  =  -\frac{\eta^2}{2} \sum_{\bm k\sigma}  d_{\bm k}^2 n^\prime(\epsilon_{\bm k}) 
- \frac{ U \eta^2 }{2} \sum_{\bm k\bm k^\prime}  
 d_{\bm k}  d_{\bm k^\prime} n^\prime (\epsilon_{\bm k})  n^\prime (\epsilon_{\bm k^\prime}). 
\label{append_eq:F_SO}
\end{aligned} \end{equation} 

Instead of expanding $F_{fl}$ in powers of $\eta$, it is convenient to expand the electronic propagators as 
\begin{equation} \begin{aligned}
G_{k \sigma} = G^0_{ k } -\sigma d_{\bm k} (G^0_{k})^2 \eta 
+ d_{\bm k}^2 (G^0_{ k})^3 \eta^2 + O(\eta^3),
\end{aligned} \end{equation} 
where $G^0_{\bm k} \equiv G_{\bm k \sigma} (\eta=0)$, before evaluating the frequency sums in
\begin{equation}
F_{fl} = -\frac{U^2}{2} \sum^\prime_{kk^\prime pp^\prime } G_{k\uparrow} G_{p\downarrow}G_{p^\prime \downarrow}G_{k^\prime \uparrow}.
\end{equation}
The prime indicates the conservation of 4-momentum, i.e. $\delta(k+p - k^\prime - p ^\prime)$.
The expanded fluctuation free energy has an $O(1)$ term, which is ignored for now, and $O(\eta^2)$ terms. 
The latter are grouped into two sets of terms: those with a prefactor $d_{\bm k} d_{\bm k^\prime}$, which are labelled as $F_{fl}^{(1)}$, and those with prefactor $d_{\bm k}^2$, labelled $F_{fl}^{(2)}$.
These are analysed separately.
The first set tidies to
\begin{equation} \begin{aligned}
F_{fl}^{(1)} 
&= - \eta^2 U^2 \sum_{k k^\prime} d_{\bm k} d_{\bm k^\prime}  (G_{ k}^0)^2  \chi_{k k^\prime}^0 (G_{ k^\prime}^0)^2 \\ 
&= - \eta^2 U^2 \sum_{\bm k \bm k^\prime} d_{\bm k} d_{\bm k^\prime} \frac{\partial \ }{\partial \epsilon_{\bm k}} \frac{\partial \ }{\partial \epsilon_{\bm k^\prime}} \\
&\Bigg(\sum_{i\omega_n i \omega_{n^\prime}} G_{\bm k}^0(i\omega_n) \chi_{\bm k \bm k^\prime}^0(i\omega_n - i\omega_{n^\prime}) G_{ \bm  k^\prime}^0(i\omega_{n^\prime}) \Bigg),
\label{append_eq:F_fl1_1}
\end{aligned} \end{equation} 
where $\chi_{k k^\prime}^0 = \sum^\prime_{pp^\prime} G^0_{p} G^0_{p^\prime }$ is the bare susceptibility of the non-ordered system.
The frequency sum is evaluated within an on-shell approximation, as described in Appendix \ref{Appendix:SC}, i.e.
\begin{equation} \begin{aligned}
\sum_{i\omega_n i \omega_{n^\prime}} &G_{\bm k}^0(i\omega_n) \chi_{\bm k \bm k^\prime}^0(i\omega_n - i\omega_{n^\prime})  G_{ \bm  k^\prime}^0(i\omega_{n^\prime}) \\
&\approx n(\epsilon_{\bm k}) \chi_{\bm k \bm k^\prime}^0(\epsilon_{\bm k} - \epsilon_{\bm k^\prime})  n(\epsilon_{\bm k^\prime}).
\end{aligned} \end{equation} 
Upon applying the energy derivatives in Eq.~\ref{append_eq:F_fl1_1},  we find that the term that dominates when $\epsilon_{\bm k, \bm k^\prime}$ are taken to zero is 
\begin{equation} \begin{aligned}
 F_{fl}^{(1)} =-\eta^2 U^2 \sum_{\bm k \bm k^\prime} d_{\bm k} d_{\bm k^\prime} n^\prime(\epsilon_{\bm k})  \chi_{\bm k \bm k^\prime}^0(0)  n^\prime(\epsilon_{\bm k^\prime}),
 \label{eq:append_F_fl1}
\end{aligned} \end{equation} 
since $n^\prime(\epsilon)$ is sharply peaked at the chemical potential.
The second set of terms tidies to
\begin{equation} \begin{aligned}
F_{fl}^{(2)}
&= 4 \eta^2 \sum_{ k} d_{\bm k}^2  (G_{k}^0)^3  \Sigma_{ k} \\ 
&= 2 \eta^2 \sum_{\bm k} d_{\bm k}^2  \frac{\partial^2 \ }{\partial\epsilon_{\bm k}^2} \bigg( \sum_{i\omega_n} G_{\bm k}^0(i\omega_n)  \Sigma_{\bm k}(i\omega_n) \bigg) \\
&\approx 2 \eta^2  \sum_{\bm k} d_{\bm k}^2  \frac{\partial^2 \ }{\partial\epsilon_{\bm k}^2} \bigg( n(\epsilon_{\bm k}) \Sigma_{\bm k }(\epsilon_{\bm k})\bigg),
\label{eq:append_F_fl2}
\end{aligned} \end{equation} 
in terms of the self-energy, $\Sigma_k = U^2 \sum_{k^\prime} G^0_{k^\prime} \chi_{k k^\prime}^0$.
The frequency sum has been evaluated in an on-shell approximation that assumes that the pole of $G_{\bm k}^0(i\omega_n)$ dominates the sum.
We now assume that the self-energy is dominated by its frequency-antisymmetric part, 
$\Sigma_{\bm k }(\epsilon_{\bm k}) \approx \frac{1}{2} \left( \Sigma_{\bm k}(\epsilon_{\bm k}) - \Sigma_{\bm k}(-\epsilon_{\bm k} )\right)$, and apply the energy derivatives.
The term that dominates when we take $\epsilon_{\bm k, \bm k^\prime} \rightarrow 0$ is 
\begin{equation} \begin{aligned}
F_{fl}^{(2)}
=  2\eta^2  \sum_{\bm k} d_{\bm k}^2 \ n^\prime (\epsilon_{\bm k}) \frac{\partial \Sigma_{\bm k}(\epsilon_{\bm k})}{\partial \epsilon_{\bm k}}\Bigg|_{\epsilon_{\bm k} = 0},
\label{append_eq:Ffl2}
\end{aligned} \end{equation} 
since $\Sigma^A(\epsilon)$, $(\Sigma^{A})^{\prime \prime}(\epsilon)$ and $n^{\prime\prime}(\epsilon)$ are all antisymmetric functions that vanish at the chemical potential.
The full free energy is assembled as $F = F_{SO} + F_{fl}^{(1)} + F_{fl}^{(2)}$, using Eqs.~\ref{append_eq:F_SO}, \ref{eq:append_F_fl1} and \ref{append_eq:Ffl2}, to give the result in Eq.~\ref{eq:nematic_expanded}.

\end{document}